\documentstyle[12pt]{article}   

\large
\newtheorem{definition}{Definition}[section]
\newtheorem{lemma}{Lemma}[section]

\newcommand{\beq}{\begin{equation}}
\newcommand{\eeq}{\end{equation}}
\newcommand{\be}{\begin{equation}}
\newcommand{\ee}{\end{equation}}
\newcommand{\ba}{\begin{eqnarray}}
\newcommand{\ea}{\end{eqnarray}}
\newcommand{\AmG}{{\cal A}/{\cal G}}
\newcommand{\AmGb}{\overline{\AmG}}
\newcommand{\Lp}{{\cal L}_p}
\newcommand{\HG}{{\cal HG}}
\newcommand{\HA}{{\cal HA}}
\makeatletter
\@addtoreset{equation}{section}
\makeatother

\def\Rl{{\mathchoice 
{\setbox0=\hbox{$\displaystyle\rm R$}\hbox{\hbox to0pt
{\kern0.4\wd0\vrule height0.9\ht0\hss}\box0}}
{\setbox0=\hbox{$\textstyle\rm R$}\hbox{\hbox to0pt
{\kern0.4\wd0\vrule height0.9\ht0\hss}\box0}}
{\setbox0=\hbox{$\scriptstyle\rm R$}\hbox{\hbox to0pt
{\kern0.4\wd0\vrule height0.9\ht0\hss}\box0}}
{\setbox0=\hbox{$\scriptscriptstyle\rm R$}\hbox{\hbox to0pt
{\kern0.4\wd0\vrule height0.9\ht0\hss}\box0}}}}
\def\Co{{\mathchoice
{\setbox0=\hbox{$\displaystyle\rm C$}\hbox{\hbox to0pt
{\kern0.4\wd0\vrule height0.9\ht0\hss}\box0}}
{\setbox0=\hbox{$\textstyle\rm C$}\hbox{\hbox to0pt
{\kern0.4\wd0\vrule height0.9\ht0\hss}\box0}}
{\setbox0=\hbox{$\scriptstyle\rm C$}\hbox{\hbox to0pt
{\kern0.4\wd0\vrule height0.9\ht0\hss}\box0}}
{\setbox0=\hbox{$\scriptscriptstyle\rm C$}\hbox{\hbox to0pt
{\kern0.4\wd0\vrule height0.9\ht0\hss}\box0}}}}

\title{Closed formula for Wilson loops for\\
$SU(N)$ Quantum Yang-Mills Theory in two dimensions}
\author{
A. Ashtekar\thanks{Center for Gravitational Physics and Geometry, 
The Pennsylvania State University, University Park, PA 16802-6300, USA}, 
J. Lewandowski\thanks{Institute of Theoretical Physics, 
University of Warsaw, 00-681 Warsaw, Poland}, 
D. Marolf\thanks{Department of Physics, The University of California, 
Santa Barbara, CA 93106, USA}, 
J. Mour\~ao\thanks{Sector de Fisica da U.C.E.H., 
Universidade do Algarve, Campus de Gambelas, 8000 Faro, Portugal}, 
T. Thiemann\thanks{Physics Department, Harvard University, Cambridge, MA
02138, USA}} 
\date{{\small Preprints CGPG-94/4-2, HUTMP-96/B-351}}

\begin{document}

\maketitle                     

\begin{abstract}

A closed expression of the Euclidean Wilson-loop functionals is
derived for pure Yang-Mills continuum theories with gauge groups
$SU(N)$ and $U(1)$ and space-time topologies
$\Rl^1\times\Rl^1$ and $\Rl^1\times S^1$. (For the $U(1)$
theory, we also consider the $S^1\times S^1$ topology.)  The treatment
is rigorous, manifestly gauge invariant, manifestly invariant under area 
preserving diffeomorphisms and handles all (piecewise analytic) loops in one
stroke. Equivalence between the resulting Euclidean theory and and the
Hamiltonian framework is then established. Finally, an extension of
the Osterwalder-Schrader axioms for gauge theories is proposed. These 
axioms are satisfied for the present model.

\end{abstract}

\section{Introduction}

Although the literature on Yang-Mills theories in 2 space-time
dimensions is quite rich, a number of issues have still remained
unresolved. The purpose of this paper is to analyze three such issues.
The paper is addressed both to high energy theorists and mathematical
physicists. Therefore, an attempt is made to bridge the two sets of
terminologies, techniques and conceptual frameworks.

The first issue concerns the expectation values of traces of
holonomies of the connection around closed loops in the Euclidean
domain, i.e., the Wilson loop functionals. The traces of holonomies
are, arguably, {\it the} central observables of the (pure) Yang-Mills
theory. In the classical regime, they constitute a natural set of
(over)complete gauge invariant functions of connections with rich
geometrical and physical content. Hence, their Euclidean vacuum
expectation values are the natural {\it gauge invariant} analogs of
the expectation values $\chi(f) := <\exp i\int d^n x \phi(x) f(x)>$ in
scalar field theories which determine all the $n$-point
(i.e. Schwinger) functions (via repeated functional differentiation
with respect to $f$.)  From theoretical physics considerations,
therefore, one expects the Wilson loop functionals to completely
determine the theory. From a mathematical physics perspective, the
quantum theory is completely determined if one specifies the
underlying measure $d\mu$ --the rigorous analog of the heuristic
expression $``\, \exp -S[A] {\cal D}A $'' -- on the space of Euclidean
paths. The expectation values of products of traces of holonomies
determine the ``moments'' of the measure $d\mu$. Hence, one expects
them to determine the measure completely.

Over the years, these considerations inspired a number of authors to
devise imaginative ways to explore properties of the Wilson loop
functionals. For example, Makeenko and Migdal \cite{migdal} formulated
differential equations that these functions have to satisfy {\it on
the space of loops} and then introduced physically motivated ans\"atze
to solve them. Similarly, Gross and co-authors \cite{1} have used
stochastic methods to obtain closed expressions for non-overlapping
Wilson loops. While these methods have yielded a wealth of insights, a
closed expression for generic Wilson loops has not yet appeared in the
literature. The first purpose of this paper is to provide such an
expression for $SU(N)$ (and $U(1)$) gauge theories assuming that the
underlying Euclidean space-time has a topology of $\Rl^1\times\Rl^1,
\mbox{ or } \Rl^1\times S^1$. (In the $U(1)$ case, we also allow the
topology to be $S^1\times S^1$.)

The second issue treated here is the relation between the Euclidean
description in terms of functional integrals and the canonical
description in terms of a Hilbert space and a Hamiltonian. For scalar
field theories, there exists a general framework that ensures this
equivalence (see, e.g. \cite{11}). We extend it to gauge theories and
explicitly establish the equivalence between the two descriptions in
the case when the Euclidean topology is $\Rl^1\times\Rl^1$ or
$\Rl^1\times S^1$.  While the extension involved is rather
straightforward, it is quite illuminating to see how the Euclidean
framework --which, a priori, does not know that the system has only a
finite number of true degrees of freedom-- reduces to the Hamiltonian
framework which, from the very beginning, exploits the fact that this
is a quantum mechanical system, disguised as a quantum field theory.

Our third goal is to suggest an extension of the axiomatic framework
of Osterwalder and Schrader. In that framework, one assumes from the
very beginning that the underlying space of paths is linear, and can
be identified with the distributional dual ${\cal S}'$ of the Schwartz
space ${\cal S}$ of smooth test functions of rapid decrease (see,
e.g., \cite{11}).  The axioms are restrictions on the measure $\mu$ on
${\cal S'}$, formulated as conditions on the functional $\chi(f) =
\int d\mu (\exp i\int d^nx \phi(x) f(x))$, introduced above, now
interpreted as the Fourier transform of the measure $\mu$. Now, in
gauge theories, it is natural to regard each gauge equivalence class
of connections as a distinct physical path. The space $\AmG$ of paths
is then a genuinely non-linear space and the standard axioms can not
even be stated unless one introduces, via gauge fixing, an artificial
linear structure on $\AmG$. (In higher dimensions, due to Gribov
ambiguities, such a gauge fixing does not exist.)  We will suggest a
possible extension of the standard framework to encompass gauge
theories in a manifestly gauge invariant fashion and show that the
axioms are in fact satisfied in the 2-dimensional Yang-Mills theories
discussed in sections 2-4.  We would like to emphasize, however, that
there is a key difference between the status of the first two sets of
results and the third. In the first two cases, we deal only with
2-dimensional Yang-Mills theory and the results are definitive. In the
third part, the general framework is applicable to gauge theories in
any space-time dimension and the discussion is open-ended; it opens a
door rather than closing one. In particular, relative to other
attempts \cite{13} in the literature, our approach is still very much
in the preliminary stage.

The main ideas behind our approach can be summarized as follows. (For
a more detailed discussion, see \cite{26,BanachProc}.) First, we will
maintain manifest gauge invariance in the sense that we will work
directly on the space $\AmG$. No attempt will be made to impose a
vector space structure by gauge-fixing; we will face the
non-linearities of $\AmG$ squarely. Now, it is well-known that, in
quantum field theory, smooth fields make a negligible contribution to
the path integrals; physically interesting measures tend to be
concentrated on distributions. Therefore, in the case of gauge
theories, we need to allow generalized connections. Fortunately, a
suitable completion $\AmGb$ of the space $\AmG$ of smooth physical
paths has been available in the literature for some time \cite{3,2}. 
Furthermore, this space carries \cite{2,29,baez} a rigorously defined, 
uniform measure $\mu_0$ which can serve as a fiducial measure  ---the
analog of the heuristic measure ${\cal D}A$. The idea is to construct
the physically relevant measure by ``multiplying $d\mu_0$ by $\exp
-S$'', where $S$ is the Yang-Mills action.

As in all constructive quantum field theories, this task is, of
course, highly non-trivial. We proceed in the following steps. First,
we consider Wilson's lattice-regularized version $S_{W}$ of $S$. Now,
it turns out that $\exp -S_{W}$ {\it is} an integrable function with
respect to the measure $d\mu_0$ and, furthermore, products of traces
of holonomies, $T_{\alpha_1}\, ...\, T_{\alpha_k}$, around loops
$\alpha_1, \,..., \,\alpha_k$ are integrable on $\AmGb$ with respect
to the measure $\exp -S_{W}\, d\mu_0$. We compute these expectation
values as a function of the lattice spacing, used in the Wilson
regularization, and then show that the resulting expressions have a
well-defined limit as the spacing goes to zero. These are the required
Wilson-loop functionals in the continuum. General theorems
\cite{3,2,29} from integration theory on $\AmGb$ guarantee that there
exists a genuine, normalized measure $\mu_{\rm YM}$ on $\AmGb$ such
that the integrals of products of traces of holonomies with respect to
$\mu_{\rm YM}$ are the Wilson loop functionals computed by the
regularization procedure. This provides a concrete proof of the
existence of a consistent Euclidean theory.

The techniques we use were first developed in the context of a
non-perturbative approach to general relativity \cite{15}.  Therefore,
our emphasis is often different from that in the literature of
Yang-Mills theories. For instance, we arrive at the final, closed
expressions of Wilson loops by a direct computation of the functional
integrals, rather than through differential equations these
functionals satisfy on the loop space. In this sense, our approach is
similar to that followed in the mathematical physics
literature. However, in these rigorous approaches, one often tries to
exploit methods which have been successful in kinematically linear
theories and, to do so, introduces a vector space structure of $\AmG$
through gauge fixing. As mentioned above, we work directly on the
non-linear space $\AmG$ and thus avoid gauge fixing in conceptual
considerations. Also, our method respects the invariance of the theory
under area preserving diffeomorphisms. In particular, our Wilson loop
functionals --and hence the final, physical measure for the continuum
theory-- are manifestly invariant under the action of this group.

The plan of the paper is as follows. In section 2, we review the
relevant notions from calculus on $\AmG$.  In section 3, we
reformulate lattice gauge theory in a manner that makes the analytic
computation of Wilson loop functionals easier.  This formulation
constitutes the basis of our discussion of the continuum theory in
section 4. Here, we first derive the general form of the Wilson loop
functionals with ultra-violet and infrared cut-offs provided by the
lattice regularization and then show that the functionals admit
well-defined limits as the cut-offs are removed. In the mathematical
physics terminology, these limits are the generating functions for the
physical, Yang-Mills measure on $\AmGb$.  For simple loops, we recover
the well-known area law which is generally taken to be the signature
of confinement. More generally, if we suitably restrict our choices of
loops, our general results reduce to those obtained previously in the
mathematical physics literature.  Section 4 reviews the Hamiltonian
quantization of Yang-Mills fields in cases when the underlying
Lorentzian space-time has the topology of a 2-plane or a
2-cylinder. The aim of section 5 is three-folds. We begin with a brief
review of the Osterwalder-Schrader framework for kinematically linear
theories and, using the machinery developed in sections 2-4, propose an
extension to handle gauge theories. We then show that our
2-dimensional model, treated in section 4, satisfies these
axioms. Finally, we show that the Hamiltonian framework reviewed in
section 4 can be systematically recovered from the Euclidean
framework. Section 6 summarizes the main results, compares them with
the results available in the literature and suggests some directions
for further work.

A number of technical topics are covered in appendices. Specifically,
Young tableaux which are needed in certain computations of section 4
are discussed in Appendix A and the details of the Euclidean $U(1)$
theory on a torus are presented in Appendix B.

\section{Preliminaries}

In this section, we will review the basic notions from
\cite{3,2,29,baez,4} (and references therein) which will be used in
this paper. This material will provide the necessary background for
our discussion of the mathematical aspects of functional integration,
axiomatic formulation of gauge theories and the relation between
Euclidean and Hamiltonian formulations.  A reader who is interested
primarily in the computation of the Wilson loop functionals can skip
this material and go directly to sections 3 and 4.

By a loop we will mean a piecewise-analytic embedding of $S^1$ into
the (Euclidean) space-time manifold $M$. For technical convenience, we
will only consider based loops, i.e., loops passing through a fixed
point $p$ in $M$. Denote the set of these loops by $\Lp$. As indicated
already, our structure group will be either $SU(N)$ (where $N\ge 2$) or
$U(1)$. Fix any one of these groups, consider a trivial Principal
fibre bundle $B$ on $M$ and denote by ${\cal A}$ the space of smooth
connections on $B$. Given any $A\in {\cal A}$, we can associate with
every $\alpha\in \Lp$ an element of $SU(N)$ by evaluating the holonomy,
\beq\label{2.1}
h_\alpha(A):={\cal P} \exp(-\oint_\alpha A)\, , 
\eeq
at the base point $p$ (where, as usual, ${\cal P}$ stands for ``path
ordered'').  Let us introduce an equivalence relation on $\Lp$: two
loops $\alpha_1,\alpha_2\in\Lp$ will be said to be holonomically
equivalent, $\alpha_1\sim\alpha_2$, iff
$h_{\alpha_1}(A)=h_{\alpha_2}(A)\,\, \forall A\in {\cal A}$. Each of
these holonomically equivalent loops will be called a {\it hoop}. It
is straightforward to verify that the space $\HG$ of hoops has a
natural group structure. We will call it the {\it Hoop group}. For
notational simplicity, in what follows we will not distinguish between
a hoop and a loop in the corresponding equivalence class.

Denote by ${\cal G}$ the group of smooth, local gauge transformations
(i.e., of smooth vertical automorphisms of $B$). Of special interest
are the ${\cal G}$-invariant functions $T_\alpha$ of connections
obtained by taking traces of holonomies:
\beq 
T_\alpha(A):=\frac{1}{N}\mbox{tr}(h_\alpha(A)) 
\eeq 
where the trace is taken with respect to the $N$-dimensional
fundamental representation of the structure group. As is well known,
the functions $T_\alpha$ suffice to separate points of $\AmG$ in the
sense that given all the $T_\alpha$, we can reconstruct the smooth
connection modulo gauge transformations \cite{5}. This is significant
because, in the classical theory, physical paths are represented by
elements of $\AmG$.
 
To go over to the quantum theory, we need to extend this space of
paths appropriately since the set of smooth paths is, typically, of
zero measure in physically interesting theories. One possible
extension has been carried out in the literature \cite{3,2}.  (For
motivational remarks, see \cite{26}.) This extension, $\AmGb$,
can be characterized in three complementary ways, each emphasizing a
different set of its properties. Since $\AmGb$ will play a fundamental
role in the quantum theory --in our approach it represents the space
of gauge invariant, physical paths in the Euclidean approach-- we will
now sketch all these characterizations:
\begin{itemize}
\item[i)] Perhaps the simplest characterization is the following:
$\AmGb$ is the space of {\it all} homomorphisms from the hoop group
$\HG$ to the structure group $SU(N)$ or $U(1)$, (modulo the adjoint
action of the structure group at the base point $p$.) It is obvious
that, given a smooth connection, the holonomy map of (\ref{2.1})
provides such a homomorphism. However, it is easy to construct
\cite{2} examples of more general homomorphisms which, for example,
would correspond to ``distributional connections''. In relation to 
the more familiar scalar field theories, $\HG$ will play a
role which in some ways is similar to that played by the space ${\cal
S}$ of test functions and $\AmGb$ is analogous to the space ${\cal S}'$
of Schwartz distribution. In particular, just as ${\cal S}'$ is the
space of paths for scalar fields, $\AmGb$ will serve as the space of
paths for gauge theories. The ``duality'' between $\HG$ and $\AmGb$ is
non-linear. However, just as elements of ${\cal S}$ serve as labels for
cylindrical functions on ${\cal S}'$, elements of $\HG$ will serve as
labels for cylindrical functions on $\AmGb$.

\item[ii)] The second characterization brings out the topological
structure of $\AmGb$. Recall first that in any of the standard Sobolev
topologies on $\AmG$, the functions $T_\alpha$ are
continuous. Furthermore, for gauge groups under consideration,
they are bounded. Hence, the $\star$-algebra they generate is a
sub-algebra of the $C^\star$-algebra $C^0(\AmG)$ of all continuous
bounded functions on $\AmG$. Denote the completion of this
$\star$-algebra by $\HA$. This is an Abelian $C^\star$-algebra with
identity and is called the {\it holonomy algebra}. Now, the Gel'fand
representation theory guarantees that $\HA$ is naturally isomorphic
with the $C^\star$-algebra of all continuous functions on a compact
Hausdorff space. This space  --the Gel'fand spectrum of $\HA$--  is our
$\AmGb$. Thus, the topology on $\AmGb$ is the coarsest one which
makes the Gel'fand transforms of the traces of holonomies
continuous. Finally, since $\HA$ suffices to separate points of
$\AmG$, it immediately follows that $\AmG$ is densely embedded in
$\AmGb$.

\item[iii)] The last characterization is in terms of projective limits
\cite{12}. One begins with two projective families labelled by graphs,
each consisting of compact Hausdorff manifolds. The projective limit
of the first yields a completion $\bar{\cal A}$ of the space ${\cal
A}$ of smooth connections while the projective limit of the second
provides a completion of the the space ${\cal G}$ of smooth gauge
transformations. One then shows that $\AmGb = \bar{\cal A}/\bar{\cal
G}$. This characterization is best suited for analyzing the
(surprisingly rich) geometric structure of $\AmGb$ \cite{29,AL}.
\end{itemize}

Finally, we note that $\AmGb$ admits \cite{2,baez,29} a natural,
normalized, Borel measure $\mu_0$ which, in our approach, will play
the role that ``${\cal D}\!A$'' plays in heuristic considerations. We
will conclude by indicating how this measure is defined.

To begin with, let us consider the family of all piecewise analytic,
oriented graphs $\Gamma$ in $M$. Denote by $\pi_1(\Gamma)$ the
fundamental group of the graph $\Gamma$ . Choose a system of
generators $\beta_1,..,\beta_n$ of $\pi_1(\Gamma)$ where
$n:= \dim(\pi_1(\Gamma))$ is the number of independent
generators of the fundamental group.  With this machinery at hand, we
can define the notion of ``cylindrical functions'', which will be the
simplest functions on $\AmGb$ that we will be able to integrate. Note
first that, given any graph $\Gamma$, we have a natural projection
map,
\beq 
p_\Gamma\; :\; \AmGb\rightarrow G^{n}\, \quad \,
A\rightarrow (h_{\beta_1}(A),..,h_{\beta_{n}}(A)) \; , 
\eeq
from $\AmGb$ to $G^{n}$, where $G$ is the structure group (i.e.
$SU(N)$ or $U(1)$) under consideration. Cylindrical functions are
obtained by pull-backs of smooth functions on $G^{n}$ under
this map. Thus, given any smooth function $f_\Gamma$ on $G^{n}$,
$f=(p_\Gamma)^*\, f_\Gamma$ is a cylindrical function.

The measure $\mu_0$ on $\AmGb$ can now be introduced via:
\beq 
\int_{\AmGb} d\mu_0(A) f(A) :=\int_{G^{n}}d\mu_H(g_1)..
d\mu_H(g_{n})\,\, f_\Gamma(g_1,..,g_{n}) \; . 
\eeq 
The proof that this condition does indeed define an infinite
dimensional, ($\sigma$-additive) regular, normalized Borel
measure $\mu_0$ on $\AmGb$ is given in Ref. \cite{4}.

\section{Lattice gauge theory}

In this section, we will recast the standard description of lattice
gauge theory in a form that is better suited for our discussion of the 
continuum limit in section 4.

Consider finite square lattices $\Gamma(a,L_x,L_y)$ in $M$ with
spacing $a$ and length $L_x \mbox{ and } L_y$ in the $x$ and $y$
directions.  This lattice contains $(N_x+1)(N_y+1)$ vertices, where
$N_x a:=L_x,\; N_y a:=L_y$.  Note that the use of such a lattice for
quantum field theory implies both an infra-red regulator (the finite
volume defined by the $L_x\mbox{ and }L_y$) and an ultra-violet
regulator (defined by the lattice spacing $a$).  Our strategy will be
to construct a regulated quantum theory in this section and then
remove the regulators in the next section.

Let us denote the open path along an edge (link) of the lattice from 
a vertex $i$ to an adjacent vertex $j$ by
$$ l=l_{i\to j} $$ so that we may
define the plaquette loops
\beq 
\Box_{(x,y)}:=l_{(x,y)\to(x,y+1)}^{-1}\circ 
l_{(x,y+1)\to(x+1,y+1)}^{-1}\circ
l_{(x+1,y)\to(x+1,y+1)}\circ l_{(x,y)\to(x+1,y)}. 
\eeq
That is, each plaquette loop starts at the bottom left corner and our
convention is such that the coordinate directions define positive
orientation. Here the coordinates $x,y$ are taken to be integers.  For
the plane $M=\Rl\times \Rl$, all of these links are distinct while for
the cylinder, $M=\Rl\times S^1$, we identify $l_{(1,y) \to (1,y+1)}
\equiv l_{(N_x+1,y)\to(N_x+1,y+1)}$. On the torus, we also identify
$l_{(x,1)\to(x+1,1)}\equiv l_{(x,N_y+1)\to(x+1,N_y+1)}$.

Next, we introduce a set of closed loops which can serve as
generators, i.e., in terms of which any loop in $\Gamma$ based at $p$
can be expressed via composition:
\begin{itemize}
\item[i)]Let $\rho_{x,y}$ be an open path in $\Gamma$ from 
$p$ to the point $(x,y)$. The loops
\beq 
\beta_{x,y}:=\beta_{\Box_{(x,y)}}:=\rho_{x,y}^{-1}\circ\Box_{(x,y)}\circ
\rho_{x,y} \; 
\eeq
generate all loops on the plane.  
\item[ii)] On the cylinder, we need an additional loop.  We will take
it to be the `horizontal' loop 
\beq
\gamma_x:=l_{(N_x,N_y+1)\to(1,N_y+1)}\circ l_{(N_x-1,N_y+1)\to
(N_x,N_y+1)} \circ...\circ l_{(1,N_y+1)\to (2,N_y+1)} \;. 
\eeq
\item[iii)] Similarly, on the torus we need an additional loop, 
\beq 
\gamma_y:=l_{(1,N_y)\to(1,1)}\circ l_{(1,N_y-1)\to (1,N_y)}
\circ\, ...\, \circ l_{(1,1)\to (1,2)} \;. 
\eeq
However, the loops $\{ \beta_{x,y}, \gamma_x, \gamma_y \}$ are not
independent\footnote{An intuitive notion of independence will suffice
for our work here.  For a careful definition, see \cite{2}.} as the
loop $\gamma_y^{-1}\circ\gamma_x^{-1}\circ\gamma_y\circ\gamma_x$ can
be written as a composition of the $\beta_{x,y}$.  This constraint
will lead to an ``interacting'' $U(1)$ theory for the torus in contrast
to the plane and the cylinder. 
\end{itemize}

With these preliminaries out of the way, let us now summarize the
standard formulation of the lattice gauge field theory by Wilson (see,
for example, \cite{8}). For each of the links in the lattice,
introduce one $G$-valued degree of freedom (the `parallel transport
along the link').  Let the `lattice Yang-Mills action' be given by the
Wilson expression
\beq 
S_{W}:=\sum_\Box [1-\frac{1}{N}\Re \mbox{tr}(h_\Box)], 
\eeq 
where $h_\Box$ denotes product of link variables around the plaquette
$\Box$ and $\Re \rm{tr}$ is the real part of the trace.  Also, let
$d\mu_W$ be the Haar measure on $G^{N_{l}}$, where $N_{l}$ is
the number of links in the graph. The regulated Wilson-loop functional 
is now given by 
\beq 
<T_{\alpha_1}..T_{\alpha_k}>
:= \frac{1}{Z(a;L_x,L_y)}\int_{G^{N_{l}}}\,  d\mu_W\,\,
e^{-\beta S_{W}}\, T_{\alpha_1}... T_{\alpha_k} 
\eeq
where $\alpha_1, ... \alpha_{k}$ are loops in $\Gamma =
\Gamma(a;L_x,L_y)$; the ``inverse temperature" is given by 
\beq
\beta=\frac{1}{g_0^2 a^{4-d}} 
\eeq 
($d=2$ being the dimension of $M$); and where $g_0=g_0(a)$ is the bare
coupling constant. The partition function $Z=Z(a;L_x,L_y)$ is defined
through $<T_p\, ...\, T_p> =1$ where $p$ denotes the trivial loop at
$p$. From a mathematical physics perspective, these Wilson loop
functionals can also be regarded as the characteristic functional of
the regulated measure. To emphasize this dual interpretation, using
the standard notation for characteristic functionals, we will set:
\be
\chi(\alpha_1,..\alpha_{k};a;L_x,L_y):= <T_{\alpha_1}..T_{\alpha_k}>\, .
\ee

For our purposes, it will turn out to be more convenient to re-express
the characteristic functional in terms of integrals over the
independent {\it loops} in the graph $\Gamma$.  To do so, we make use
of the fact that, whenever it is used to integrate gauge invariant
functions, the measure $d\mu_W$ may be replaced by the Haar measure on
$G^{N_{ig}}$, where $N_{ig}$ is the number of independent loop
generators of the graph $\Gamma$.  This fact follows immediately from
the results of \cite{baez,AL}.  (In the language of these works,
it is contained in the statement that $\overline{\cal A}/
\overline{\cal G} = \overline {{\cal A}/{\cal G}}$ and that the Haar
measure on $\overline{\cal A}$ projects unambiguously to yield the
Haar measure on $\overline{{\cal A} / {\cal G}}$.)  Thus, we may write
the regulated characteristic functional as: 
\ba 
\chi(\alpha_1,..\alpha_k)=\frac{1}{Z} \int_{G^{N_{ig}}} \prod_\Box
d\mu_H(g_\Box)& & \exp(-\beta S_{W}) \times\\ 
& & \times \left\{\begin{array}{ll} \prod_{i=1}^k \mbox{tr}\ 
\alpha_i(g_\Box) &\mbox{:on } \Rl^2\\ 
\int_G d\mu_H(g_x) \prod_{i=1}^k \mbox{tr} \
\alpha_i(g_\Box, g_x) & \mbox{: on } \Rl^1\times S^1 \end{array}
\right. \nonumber 
\ea 
where $d\mu_H$ is the Haar measure on $G$ and $\alpha_i(g_\Box)$ is
the expression for $\alpha_i$ in terms of the generators $\beta_{x,y}$
with each generator $\beta_{x,y}$ replaced by the integration variable
$g_{x,y}$ (similarly for $\alpha_i(g_\Box, g_x)$).  The corresponding
expression for the torus will appear at the end of this section.  The
idea of the next section will simply be to evaluate the above
integrals for any given $a, L_x,L_y$ and then take the limits to
remove the ultra-violet and infra-red regulators.

To conclude this section, we will introduce some definitions
and collect a few facts about loops in $\Gamma$. These will be useful
in section 4.

\begin{definition}
A loop is said to be simple iff there is a holonomically equivalent
loop which has no self-intersections.
\end{definition}
Note that any simple homotopically trivial loop divides the space-time
into two regions: an interior which is topologically a 2-disk and an
exterior.  This is just the Jordan curve theorem.
\begin{definition}
On the torus, we define the surface enclosed by a simple homotopically
trivial loop to lie on the left as one follows the loop
counterclockwise (when the torus is represented as a 2-dimensional
rectangle with the standard identifications.)
\end{definition}
\begin{definition}
Two distinct simple homotopically trivial loops are said to be
non-over\-lapp\-ing iff the intersection of the surfaces that they enclose
has zero area.  The homotopically non-trivial loops $\beta_x$ and
$\beta_y$ will both be said not to overlap any other
loop. 
\end{definition} 
So, for example, all the loops $\beta_{x,y}$ are simple since they lie
in the same hoop class as the plaquette loops $\Box_{(x,y)}$.
Non-overlapping distinct simple loops may share whole segments whence
the plaquette generators of our graph (lattice) are mutually
non-overlapping.

It will turn out that the following two simple lemmas govern the form
of the characteristic functional in two space-time dimensions.
\begin{lemma}
Every simple, homotopically trivial loop $\alpha$ on $\Gamma$ can be written 
as a particular composition of the generators $\beta_\Box$ 
contained in the surface enclosed by $\alpha$, with each
$\beta_\Box$ appearing once and only once. 
\end{lemma}
It is readily checked that when two homotopically trivial loops
$\alpha_1$ and $\alpha_2$ (enclosing disks $D_1$ and $D_2$) are
non-overlapping and such that $D_1 \cup D_2$ is also a disk, then
either $\alpha_1 \alpha_2$ or $\alpha_1 \alpha_2^{-1}$ (or, on the
torus, perhaps the inverse of one of these loops) encloses $D_1 \cup
D_2$.  Since every disk is a finite union of plaquettes, Lemma 3.1
follows immediately. $\Box$

This Lemma allows us to write a simple expression for the generating
functional on the torus.  Note that, after `ungluing' the torus to
make a rectangle, the loop $\gamma_x \circ \gamma_y \circ \gamma_x^{-1}
\circ \gamma_y^{-1}$ is simple and homotopically trivial, enclosing
the entire area of the torus.  As a result, it may be written as a
product of the plaquette loops $\beta_\Box$ in which each $\beta_\Box$
appears once and only once.  We may therefore pick any one of these
loops (say $\beta_{(0,0)}$) and write it as a function of the other
plaquette loops and the loops $\gamma_x, \gamma_y$. Alternatively, we
find a product $C$ of holonomies along all the loops
$\beta_\Box,\gamma_x, \gamma_y$ which is the identity in
$G$. Inserting a $\delta$ distribution on $G$ enforcing the constraint
$C=1_N$ we find for the generating functional on the torus
\ba
\chi(\alpha_1,..\alpha_k)&=&\frac{1}{Z} \int_{G^{N_{ig}}} 
\prod_{\Box}d\mu_H(g_\Box) d\mu_H(g_x)d\mu_H(g_y)\exp(-\beta S_{W}) 
\prod_{i=1}^k \mbox{tr}\ \alpha_i(g_\Box,g_x,g_y) \cr
&=&\frac{1}{Z} \int_G \prod_\Box
d\mu_H(g_\Box) d\mu_H(g_x)
d\mu_H(g_y) \delta(C,1_N)\exp(-\beta S_{W}) \cr
& \times & \prod_{i=1}^k \mbox{tr}\ \alpha_i(g_\Box,g_x,g_y) \;.
\ea

Finally, we have:
\begin{lemma}
Every loop can be written as a composition of simple non-overlapping
loops.
\end{lemma}
This follows from the fact that the $\beta_\Box$ (together
with $\beta_x,\beta_y$ on $S^1 \times \Rl$ and $T^2$) are simple
and non-overlapping and that they generate the graph $\Gamma$. $\Box$

\section{Continuum theory}

In this section we will derive a closed expression for the Wilson loop
functionals --i.e., for the characteristic functional of the measure--
for the continuum theory when the underlying manifold $M$ is either a
2-plane or a cylinder. (For the torus, we have been able to carry out
the computation to completion only for the Abelian case, $G= U(1)$,
and this theory is discussed in detail in Appendix B.)

In section 4.1, we will discuss $U(1)$ theories and in section 4.2,
$SU(N)$ theories. In both cases, we will show that the
lattice-regulated characteristic functional admits a well-defined
limit as the ultra-violet and infrared cut-offs are
removed. Furthermore, we will be able to read-off certain qualitative
properties of these functionals. However, the explicit expression
involves a group-dependent constant. This is evaluated in section 4.3.

\subsection{Abelian case $(U(1))$}
 
Let us first note that, in the $U(1)$ case, products of functions
$T_\alpha$ can be reduced to a single $T_{\alpha'}$ in the obvious
fashion. Therefore, we need to consider only single loops. Fix a loop
$\alpha$ and consider its decomposition in to non-overlapping simple
loops. Let $k_I$ be the effective winding number of the simple
homotopically trivial loop $\alpha_I,\; I=1,..,,n$ and let $l_x, l_y$
be winding numbers of the homotopically non-trivial loops
$\beta_x,\beta_y$ in this decomposition. Define $|\alpha_I|$ to be the
number of plaquettes enclosed by the simple loop $\alpha_I$. We can
then write the characteristic functional as follows (with $G=U(1)$):
\begin{eqnarray*} 
\chi(\alpha) & = & \frac{1}{Z}\int\prod_\Box[\int_G d\mu_H(g_\Box)
\exp(-\beta(1-\Re(g_\Box)))
\prod_{I=1}^n (\prod_{\Box\in\alpha_I}g_\Box)^{k_I}]\, \times 
\nonumber\\
& & \times\, \int_G d\mu_H(g_x) g_x^{l_x}\int_G d\mu_H(g_y) 
g_y^{l_y} \delta(\prod_\Box g_\Box,1)^e\, ,
\end{eqnarray*}
where we could neglect the precise ordering of plaquette variables
(that occurred in the decomposition of $\alpha$ in terms of
$\beta_\Box,\beta_x, \beta_y$) because the gauge group is Abelian. In
this formula $l_x=l_y=0$ on the plane and $l_y=0$ on the cylinder and
$e=0$ for the plane and the cylinder while $e=1$ for the torus. Now,
for $G=U(1)$, we have $\int_G d\mu_H(g)g^n=\delta(n,0)$. Hence, it
follows immediately that the characteristic functional is non-zero if
and only if $l_x=l_y=0$.  Therefore, we will focus on this case in the
sequel.

Now, let us consider the partition function, $Z$. For the plane and the
cylinder, different plaquette contributions decouple and we obtain: 
\beq 
Z=[\int_G d\mu_H(g)\exp(1-\Re(g))]^{N_x N_y}
\eeq 
For the torus, on the hand, decoupling does not occur and we are left 
with
\beq 
Z=\int\prod_\Box d\mu_H(g_\Box)\exp(-(1-\Re(g_\Box)))
\delta(\prod_\Box g_\Box,1_N)\, . 
\eeq 
Thus, even in the Abelian, $U(1)$ case, the Euclidean theory in two
space-time dimensions has interactions! We will continue the discussion
of this case in Appendix B.

Collecting these results, for the plane and the cylinder, we can now
reduce the expression of $\chi(\alpha)$ to:
\beq \label{chi}
\chi(\alpha)=\prod_{I=1}^n \, [\frac{\int_G d\mu_H(g)
\exp(-\beta(1-\Re(g))) g^{k_I}} {\int_G
d\mu_H(g)\exp(-\beta(1-\Re(g)))}]^{|\alpha_I|} 
\eeq 
in case when $l_x=0 $ (and $\chi(\alpha) = 0$ otherwise.) We now
want to take the continuum limit. The ultra-violet limit corresponds to
letting lattice spacing go to zero, i.e., $\beta\to\infty$, and the
infrared limit corresponds to letting the lattice size go to infinity,
i.e., $L_x \to \infty$ and $L_y\to\infty$.

Let us set
\beq
J_n(\beta):=\int_G d\mu_H(g)\exp(-\beta(1-\Re(g))) g^n.
\eeq
Now, since $g$ is simply a complex number of modulus one it is obvious
that the fraction $J_n(\beta)/J_0(\beta)$ in (\ref{chi}) is a real
number of modulus lower than or equal to one. Now observe that
$|\alpha_I|=g_0^2\beta A(\alpha_I)$ where $A(\alpha_I)$ is the
Euclidean area enclosed by $\alpha_I$. In the limit, $\beta\to\infty$,
the integrand of both numerator and denominator become concentrated at
$g=1$, whence we have an expansion of the form
$J_n/J_0=(1-c(1,n)/\beta)(1+0(1/\beta^2))$, where $c$ is positive because
$J_n/J_0$ approaches the value 1 from below.  Thus, it is easy to see
that 
\be \label{4.5}
\lim_{\beta\to\infty}\chi(\alpha)=\exp(-g_0^2\sum_{I=1}^n
c(1,k_I)A(\alpha_I)) 
\ee 
for $l_x=0$ and zero otherwise. We will calculate the coefficients
$c(1,n)$ in section 4.3. Finally, note that the infra-red limit is
trivial since $\chi(\alpha)$ is independent of $L_x, L_y$, (assuming
of course that they are large enough for the region under
consideration to contain the loop).

To summarize, we can arrive at the continuum characteristic functions
as follows. Given any piecewise analytic loop $\alpha'$ in $M$, we
first consider a sufficiently fine and sufficiently large lattice and
approximate $\alpha'$ by a loop $\alpha$ lying in the lattice. Then,
we express $\alpha$ as a product of non-overlapping simple loops and
compute the regulated characteristic function $\chi(\alpha)$ directly.
Finally, we take continuum limit to arrive at the final expression
(\ref{4.5}).

\subsection{Non-Abelian case $(SU(N))$}

Let us now consider the technically more difficult non-Abelian case.
As indicated before, in this discussion, we will restrict ourselves to
the plane and the cylinder.  

For $SU(N)$, the trace identities only enable one to express traces of
products of matrices as linear combinations of traces of products of
$r:=N-1$ or fewer matrices. Hence, unlike in the Abelian case, the
product $T_{\alpha_1} ... T_{\alpha_n}$ can not be reduced to a single
$T_\alpha$; we can no longer confine ourselves to single loops. Fix a
multi-loop --i.e., a set of $r$ loops-- $\alpha_1,..,\alpha_r$ and
consider its decomposition into simple, non-overlapping loops. Suppose
that, in this decomposition, there are $n$ homotopically trivial loops
$\hat{\alpha}_I$ and $c$ homotopically nontrivial loops $\gamma_i$
(clearly, $c=0$ or $c=1$). Let $|\hat{\alpha}_I|$ be the number of
plaquettes enclosed by $\hat{\alpha}_I$ and let $k_I^\pm\mbox{ and
}l_i^\pm$ be the number of times that $\hat{\alpha}_I\mbox{ and
}\gamma_i$ occur (respectively) with positive or negative power in
this decomposition. Thus, altogether, there are $b=\sum_{I=1}^n[k_I^+ +
k_I^-]+\sum_{i=1}^m[l_i^+ + l_i^-]$ factors of holonomies around the
$\hat{\alpha}_I,\;\gamma_i$ and their inverses involved in the
expansion of the product $T_{\alpha_1}..T_{\alpha_r}$. These may occur
in arbitrary order, depending on the specific loops
$\alpha_i,\;i=1,..,r$.

It is then easy to see that we can now write $T_{\alpha_1}\, ...\,
T_{\alpha_r}$ explicitly as a product of matrices representing
holonomies around simple loops, with an appropriate contraction of
matrix-indices:
\begin{eqnarray} \label{prod}
& & N^r T_{\alpha_1}\, ... \, T_{\alpha_r} \nonumber\\& = & 
\prod_{I=1}^n[\prod_{\mu=1}^{k_I^+}
(h_{\hat{\alpha}_I})^{A^{I+}_\mu}_{B^{I+}_\mu}
\prod_{\mu=1}^{k_I^-}
(h_{\hat{\alpha}_I}^{-1})^{A^{I-}_\mu}_{B^{I-}_\mu}]\nonumber \\
& &
\prod_{i=1}^c[\prod_{\nu=1}^{l_i^+}
(h_{\gamma_i})^{C^{i+}_\nu}_{D^{i+}_\nu}\prod_{\nu=1}^{l_i^-}
(h_{\gamma_i}^{-1})^{C^{i-}_\nu}_{D^{i-}_\nu}]\nonumber \\
& & 
\prod_{k=1}^b \delta^{E_k}_{F_{\pi(k)}} 
\end{eqnarray}
Here, we have the following relation between indices that are being
contracted:
\begin{eqnarray}
(E_1,..,E_b) &\equiv& (A^{1+}_1,..,A^{1+}_{k_1^+},A^{2+}_1,..,
   A^{2+}_{k_2^+},.., A^{n+}_1,..,A^{n+}_{k_n^+}, \nonumber\\
&& A^{1-}_1,..,A^{1-}_{k_1^-},A^{2-}_1,..,A^{2-}_{k_2^-},..,
   A^{n-}_1,..,A^{n-}_{k_n^-}, \nonumber\\
&&  C^{1+}_1,..,C^{1+}_{l_1^+},C^{2+}_1,..,C^{2+}_{l_2^+},..,
   C^{n+}_1,..,C^{c+}_{l_n^+}, \nonumber\\
&&  C^{1-}_1,..,C^{1-}_{l_1^-},C^{2-}_1,..,C^{2-}_{l_2^-},..,
   C^{n-}_1,..,C^{c-}_{l_n^-})\, ,
\end{eqnarray}
and similarly with the exchanges $E\leftrightarrow F, A\leftrightarrow
B, C\leftrightarrow D$; and $\pi$ is an element of the symmetric group
of $b$ elements that depends on the loops $\alpha_i$ and defines the
specific contraction involved in $T_{\alpha_1}\, ,..\, T_{\alpha_r}$.

To evaluate the expectation values of this product of traces of
holonomies, we need to expand out the inverses of matrices that appear
in (\ref{prod}) explicitly. This can be done easily using the fact
that the matrices in question are all uni-modular. We have:
\begin{eqnarray*}
(h_{\hat{\alpha}_I}^{-1})^{A^{I-}_\mu}_{B^{I-}_\mu} 
& = & \frac{1}{(N-1)!}
\epsilon^{A^{I-}_\mu E^I_{\mu,1} .. E^I_{\mu,N-1}}
\epsilon^{B^{I-}_\mu F^I_{\mu,1} .. F^I_{\mu,N-1}}\\ &&
(h_{\hat{\alpha}_I})^{F^I_{\mu,1}}_{E^I_{\mu,1}} ..
(h_{\hat{\alpha}_I})^{F^I_{\mu,N-1}}_{E^I_{\mu,N-1}}\\ & =: & 
{\cal E}^{A^{I-}_\mu E^I_{\mu,1} .. E^I_{\mu,N-1}}
_{B^{I-}_\mu F^I_{\mu,1} .. F^I_{\mu,N-1}}
(h_{\hat{\alpha}_I})^{F^I_{\mu,1}}_{E^I_{\mu,1}} ..
(h_{\hat{\alpha}_I})^{F^I_{\mu,N-1}}_{E^I_{\mu,N-1}}\, , 
\end{eqnarray*}
and similarly for the inverse of $h_{\gamma_i}$.  Finally, if we
define $n_I:=k_I^+ +(N-1)k_I^-,\;c_i:=l_i^+ +(N-1)l_i^-$ we can
rewrite (\ref{prod}) using a tensor-product notation as:
\begin{eqnarray}
& & N^r T_{\alpha_i}\, ... \, T_{\alpha_r}  \nonumber\\ 
&=&\prod_{I=1}^n (\otimes^{n_I} h_{\alpha_I})^{A^{I+}_1 ... 
A^{I+}_{k_I^+} F^I_{1,1} ... F^I_{1,N-1} ... F^I_{k_I^-,1}...
F^I_{k_I^-,N-1}}_{B^{I+}_1 ... B^{I+}_{k_I^+}
E^I_{1,1} .. E^I_{1,N-1} ... E^I_{k_I^-,1}...E^I_{k_I^-,N-1}} 
\nonumber\\ 
&&\prod_{i=1}^c (\otimes^{c_i} h_{\gamma_i})^{C^{i+}_1 ... 
C^{i+}_{l_i^+} H^i_{1,1} ... H^i_{1,N-1} ... H^i_{l_i^-,1}
...H^i_{l_i^-,N-1}}_{D^{i+}_1 ... D^{i+}_{l_i^+}
G^i_{1,1} ... G^i_{1,N-1} ... G^i_{l_i^-,1}...G^i_{l_i^-,N-1}} 
\nonumber\\ &&
\prod_{k=1}^b \delta^{E_k}_{F_{\pi(k)}} 
\prod_{I=1}^n \prod_{\mu=1}^{k_I^-} 
{\cal E}^{A^{I-}_\mu E^I_{\mu,1} ...
E^I_{\mu,N-1}}_{B^{I-}_\mu F^I_{\mu,1} ... F^I_{\mu,N-1}}
\prod_{i=1}^c \prod_{\nu=1}^{l_i^-} {\cal E}^{C^{i-}_\nu G^i_{\nu,1} ...
G^i_{\nu,N-1}}_{D^{i,-}_\nu H^i_{\nu,1} ... H^i_{\nu,N-1}} .
\end{eqnarray}
Next, let us examine the contributions from homotopically trivial
loops. Choose $\eta:=\hat{\alpha}_I$ for some I and consider the
expression 
\be (\otimes^n h_\eta)^{A_1 .. A_n}_{B_1 .. B_n}. \ee 
Label the plaquette loops enclosed by $\eta$ from $1$ to $|\eta|:=m$;
thus $h_\eta=g_1 ... g_m$, where $g_k:=h_{\Box_k}$. Then the above
expression becomes
\begin{eqnarray} \label{tp}
& & [(g_1)^{A_1}_{C_{1,1}} (g_2)^{C_{1,1}}_{C_{1,2}} .. 
(g_m)^{C_{1,m-1}}_{B_1}] ...
[(g_1)^{A_n}_{C_{n,1}} (g_2)^{C_{n,1}}_{C_{n,2}} 
... (g_m)^{C_{n,m-1}}_{B_n}]\nonumber \\
& = & [\otimes^n g_1]^{A_1 ... A_2}_{C_{1,1} ... C_{n,1}}
[\otimes^n g_2]^{C_{1,1} ... C_{n,1}}_{C_{1,2} ... C_{n,2}} ...
[\otimes^n g_m]^{C_{1,m-1} ... C_{n,m-1}}_{B_1 ... B_n}
 \nonumber \\ 
& = & ([\otimes^n g_1] [\otimes^n g_2] ... 
[\otimes^n g_m])^{A_1 ... A_n}_{B_1 ... B_n} 
\end{eqnarray}
where, in the last step we have used the product rule for tensor
products of matrices.

With these explicit expressions at hand, we can now consider the
functional integral which yields the Wilson loop functionals. In this
evaluation, each of the n-fold tensor products in (\ref{tp}) has to be
integrated with the measure 
\be d\mu(g)=d\mu_H(g)\exp(\beta/N \Re\mbox{tr}(g))\, .\ee
To carry out this task, we will use the representation theory reviewed
in appendix A.

According to appendix A, we have:
\begin{eqnarray}\label{3.10}
&& \int_G d\mu(g) \otimes^n g \nonumber\\
& = & \oplus_{\{m\}} \oplus_{i=1}^{f^{(n)}_{\{m\}}}
\int_G d\mu(g) [p^{(n)}_{\{m\},i} \otimes^n g] \nonumber\\
& = & \oplus_{\{m\}} \oplus_{i=1}^{f^{(n)}_{\{m\}}}
[p^{(n)}_{\{m\},i} \otimes^n 1_N] \frac{1}{d_{\{m\}}}
\int_G d\mu(g) \mbox{tr}([p^{(n)}_{\{m\},i}\otimes^n g]) \nonumber\\
& = & \oplus_{\{m\}} [\oplus_{i=1}^{f^{(n)}_{\{m\}}}
[p^{(n)}_{\{m\},i} \otimes^n 1_N]] \frac{1}{d_{\{m\}}}
\int_G d\mu(g) \chi_{\{m\}}(g) \nonumber\\
& = & \oplus_{\{m\}} 
[p^{(n)}_{\{m\}} \otimes^n 1_N] \frac{1}{d_{\{m\}}}
\int_G d\mu(g) \chi_{\{m\}}(g) \nonumber\\
& = & \oplus_{\{m\}} 
[p^{(n)}_{\{m\}} \otimes^n 1_N] J_{\{m\}}(\beta,N)\, .
\end{eqnarray} 
Here, in the first step, we have decomposed the matrix $\otimes^n\, g$
into a direct sum of irreducible representations, with $i$ labeling
the orthogonal equivalent representations and $m$ labeling the
equivalence classes of inequivalent representations, and $p^{(n)}_{\{ m
\},i}$ are the Young symmetrizers; in the third step, we have used
the fact that the trace is a class function ($\chi_{\{m\}}$
being the character of the representation $\{m\}$); and, in the last
step we have simply defined
\beq \label{jm}
J_{\{m\}}(\beta,N) :=\frac{1}{d_{\{m\}}}\int_G d\mu(g) 
\chi_{\{m\}}(g)\;. 
\eeq 
Finally, using the orthogonality of the projectors $p^{(n)}_{\{m\}}
\otimes^n 1_N$ we find that the integral over (4.9) becomes 
\beq
(\sum_{\{m\}} [p^{(n)}_{\{m\}} \otimes^n 1_N] [J_{\{m\}}
(\beta,N)]^{|\eta|})^{A_1 ... A_n}_{B_1 ... B_n} \; . 
\eeq 
The integral over the homotopically non-trivial loops is quite
similar, the main difference being that the measure there is the Haar
measure and that each of these loops involves just a single
integration variable. According to appendix A we find that the
integral over $\otimes^n g$ with the Haar measure is given by
\[ [p^{(n)}_0 \otimes^n 1_N] \]
where $p^{(n)}_0 $ is the projector on the trivial representation.

Collecting these results, we can write the vacuum expectation value of
$T_{\alpha_1},\, ...\, T_{\alpha_r}$ as follows. Set 
\beq J_0(N,\beta):=\int_G d\mu(g)\, . \eeq
Then,
\begin{eqnarray}\label{gf}
N^r \chi(\alpha_1,..,\alpha_r) &=&
\prod_{I=1}^n (\sum_{\{m\}}\, [\frac{J_{\{m\}}}{J_0}]^{|\hat\alpha_I|}\,
[p^{(n_I)}_{\{m\}} \otimes^{n_I} 1_N] )^{A^{I+}_1 .. A^{I+}_{k_I^+}
F^I_{1,1} .. F^I_{1,N-1} .. F^I_{k_I^-,1}..F^I_{k_I^-,N-1}}
_{B^{I+}_1 .. B^{I+}_{k_I^+}
E^I_{1,1} .. E^I_{1,N-1} .. E^I_{k_I^-,1}..E^I_{k_I^-,N-1}} \nonumber\\ &&
\prod_{i=1}^m ([p^{(m_i)}_0 \otimes^{m_i} 1_N])^{C^{I+}_1 .. C^{i+}{l_i^+}
H^i_{1,1} .. H^i_{1,N-1} .. H^i_{l_i^-,1}..H^i_{l_i^-,N-1}}
_{D^{i+}_1 .. D^{i+}_{l_i^+}
G^i_{1,1} .. G^i_{1,N-1} .. G^i_{l_i^-,1}..G^i_{l_i^-,N-1}} \nonumber\\ &&
\prod_{k=1}^b \delta^{E_1}_{F_{\pi(k)}}
\prod_{I=1}^n \prod_{\mu=1}^{k_I^-} {\cal E}^{A^{I-}_\mu E^I_{\mu,1} ..
E^I_{\mu,N-1}}_{B^{I-}_\mu F^I_{\mu,1} .. F^I_{\mu,N-1}}
\prod_{i=1}^m \prod_{\nu=1}^{l_i^-} {\cal E}^{C^{i-}_\nu G^i_{\nu,1} ..
G^i_{\nu,N-1}}_{D^{i,-}_\nu H^i_{\nu,1} .. H^i_{\nu,N-1}} .
\end{eqnarray}
This is the closed expression for the regulated Wilson loops.
Although it seems complicated at first, its structural form is rather
simple\footnote{A more elegant derivation of (\ref{gf}) uses the notion 
of a loop-network state \cite{BanachProc}, however, since products of 
traces of the holonomy are more familiar to gauge theorists we have 
refrained from introducing the associated mathematical apparatus here}. 
First of all, the lattice spacing and the coupling constant
enter this expression only through $J_{\{m\}}$. The rest is all an
explicit contraction of indices of a product of a {\it finite} number
of matrices. For any given group $G=SU(N)$, the matrices depend only
on the decomposition of $T_{\alpha_1}, ...\, T_{\alpha_r}$ in terms of
the $n$ holonomies around the homotopically trivial, simple loops and
the $m$ holonomies around the homotopically non-trivial simple loops.

To establish the existence of the continuum limit, therefore, we only
need to show that $[J_{\{m\}}(\beta)/J_0(\beta)]^{|\alpha_I|}$
converges to a finite value as $a\to 0$. Let us begin by noting that
\[ |J_{\{m\}}(\beta,N)|\le \int_G d\mu(g) |\frac{\chi_{\{m\}}(g)}
{d^{(n)}_{\{m\}}}| \le J_0(\beta,N)\, . \]
This estimate implies that $|J_{\{m\}}/J_0|$ is always a number between
$0$ and $1$ for finite $\beta$. Moreover, we have
\[ \lim_{\beta\to\infty} \frac{J_{\{m\}}}{J_0}=\lim_{\beta\to\infty}
\frac{\int_G d\mu_H\, \exp(-\beta (1-1/N\Re\mbox{tr}(g)))\,\, 
\frac{\mbox{tr}(\pi(g))}
{\mbox{dim}(\pi)}}{\int_G d\mu_H \,\exp(-\beta(1-1/N\Re
\mbox{tr}(g)))}=1  \]
since for $\beta\to\infty$ the measure in both numerator and
denominator becomes concentrated at the identity for which both
integrand are equal to the number one. Therefore, we have an
asymptotic expansion of the form
\[ \frac{J_{\{m\}}(\beta,N)}{J_0(\beta,N)}=(1-\frac{c(N,\{m\})}
{\beta}) (1+0(1/\beta^2))\, , \]
where the first order coefficient $c(N,\{m\})$ must be {\em
non-negative} since $J_{\{m\}}/J_0$ approaches unity {\em from below}.
Finally, observing that $|\alpha_I|=\beta g_0^2 A(\alpha_I)$, we find
that the continuum limit of (\ref{gf}) is given by replacing the
$[J_{\{m\}}/J_0]^{|\alpha_I|}$ by
\beq
\lim_{\beta\to\infty} [ \frac{J_{\{m\}}(\beta,N)} 
{J_0(\beta,N)} ]^{|\alpha_I|} =  
\exp(-c(N,\{m\})g_0^2 A(\alpha_I)). 
\eeq 
This establishes the existence of the continuum limit. To obtain the
explicit formula for the Wilson loops, it only remains to evaluate the
constants $c(N,\{m\})$. We will carry out this task in the next
sub-section.

We will conclude this sub-section with a few remarks. (Some of these
observations have been made in the context of other approaches but are
included here for completeness.)
\begin{itemize}
\item[i)] Although the explicit expression of the Wilson loop
functionals (or the characteristic functional for the Yang-Mills
measure on $\AmGb$) is rather complicated
\footnote{Note however that the computation only involves
complicated traces and can be performed by algebraic manipulation
programs very quickly.}, 
some of the qualitative features can be easily read-out. Note first
that if we have a single, simple loop $\alpha_0$, the expectation
value collapses to simply:
\be 
<T_{\alpha_0}> \equiv \chi(\alpha_0) = e^{-c g_0^2 A(\alpha_0)}\, , 
\ee
where $c$ is the value of the first $SU(N)$ Casimir on its fundamental
representation and where $A(\alpha_0)$ is the Euclidean area enclosed
by the loop $\alpha_0$. Thus, the area law --generally taken to be the
signal of confinement-- holds. Note that the loop does not have to be
large; the expression is exact. Finally, note from section 4.1 that
this law holds also for the {\it Abelian} theory. Thus, the continuum
limit of the lattice $U(1)$ theory provides us the confined phase of
the theory which is different from the phase described by the standard
Fock representation.
\item[ii)] More generally, if one restricts oneself to {\it
non-overlapping} loops, our expression reduces to that found by Gross
et al \cite{1}.  
\item[iii)] Note that, as in the Abelian theory, the infra-red limit
is trivial since the continuum expression of the Wilson loop
functionals does not depend on $L_x$ or $L_y$ at all (provided of course
the lattice is chosen large enough to encompass the given $r$ loops).
\item[iv)] It is interesting to note that we did not have to
renormalize the bare coupling constant $g_0$ in the process of taking
the continuum limit. This is a peculiarity of two dimensions. Indeed,
in higher dimensions, the bare coupling constant does not have the
correct physical dimensions to allow for an area law which suggests
that renormalization would be essential. 
\item[v)] In the classical theory in higher dimensions, the Yang-Mills
action depends on the space-time metric and is thus invariant only
under the action of the finite dimensional isometry group of the
underlying space-time (the Poincar\'e (respectively, Euclidean) group,
if the space-time is globally Minkowskian (Euclidean).) In two
space-time dimensions, on the other hand, one needs only an area
element to write the Yang-Mills action. Thus, the symmetry group is
considerably enlarged; it is the {\it infinite} dimensional group of
area preserving diffeomorphisms. A natural question is whether the
Wilson loop functionals are also invariant under this larger group.
Our explicit expression makes it obvious that it is. Thus, the
infinite-dimensional symmetry is carried over in-tact to the quantum
theory. This property is not obvious in many other approaches which
use gauge-fixing to endow $\AmG$ a vector space structure and then
employ the standard (space-time metric dependent) Gaussian measures in
the intermediate steps. In these approaches, special and somewhat
elaborate calculations are needed to verify invariance under all area
preserving diffeomorphisms.
\end{itemize}

\subsection{Determination of the coefficients $c(N,\{m\})$}

The main idea behind the calculation is the following; Since for
$\beta\to\infty$, the integrand of $J_{\{m\}}(\beta)$ is concentrated
at the identity, it is sufficient to calculate the integrand in Eq
(\ref{jm}) (defining $J_{\{m\}}$) in a neighborhood of the identity.

To that effect, write $g=e^A\mbox{ where }A=t^I\tau_I\in L(G)$ is in
the Lie algebra of $G$ and $t^I$ are real parameters in a neighborhood
of zero. We thus have upon inserting $g=1_N+A+\frac{1}{2}A^2+o(A^3)$
\be 
1-\frac{1}{N}\Re\mbox{tr}(g)=-\frac{1}{2N}\mbox{tr}(A^2)+o(A^3)
=\frac{1}{2}\sum_{I=1}^{\dim(G)}(t^I)^2+o(A^3) 
\ee 
where the term of first order in $A$ vanishes because it is either
purely imaginary (the Abelian sub-ideal of $L(G)$) or trace-free (the
semi-simple sub-ideal of $L(G)$) and where we have used the
normalization tr$(\tau_I\tau_J)=-N\delta_{IJ}$.  Similarly, we have an
expansion for the $\{m\}$th irreducible representation of $G$ given by
$\pi_{\{m\}}(g)=1_{\{m\}}+X+\frac{1}{2}X^2+o(X^3) \mbox{ where } X=t^I
X_I$ is the representation of the Lie algebra element $A$ in the
$\{m\}$-th irreducible representation. Then we have 
\be
\chi_{\{m\}}(g)=d_{\{m\}}+t^I\mbox{tr}(X_I)+\frac{1}{2} t^I
t^J\mbox{tr}(X_I X_J) +o(X^3) \;.
\ee 
Now, according to the Baker-Campbell-Hausdorff formula \cite{8} we
have:
\be e^{t^I\tau_I}e^{s^I\tau_I}=e^{r^I(s,t)\tau_I},\mbox{ where }
r^I(s,t)=s^I+t^I-\frac{1}{2}f^I\;_{JK}s^J
t^K+o(s^2,t^2,s^3,s^2t,st^2,t^3) 
\ee
and where $f^I_{JK}$ are the structure constants of the semi-simple
sub-ideal of $L(G)$ which therefore are completely skew. Finally, the Haar
measure can be written \cite{8} 
\be
d\mu_H(e^{t^I\tau_I})=\frac{d^{\dim(G)}t}{\det(\frac{\partial
r^I(s,t)} {\partial s^J})_{s=0}}=\frac{d^{\dim(G)}t}{1+o(t^2)} 
\ee
since $\det(\partial r/\partial s)_{s=0}=\det(1+\frac{1}{2} t^I R_I
+o(t^2))=1+\frac{1}{2}\mbox{tr} (t^I R_I)+o(t^2)=1+o(t^2)\mbox{ where
}(R_I)^J_K=f^J\;_{IK}$ is the I-th basis vector of the semi-simple
sub-ideal of $L(G)$ in the adjoint representation which is
trace-free.

We are now ready to carry out the required estimate. There exists
a subset $U_0\subset \Rl^{\dim(G)}$ which is in one-to-one
correspondence with G via the exponential map. Let $U$ be the closure 
of $U_0$ in $\Rl^{\dim(G)}$. The set $U$ is compact in $\Rl^{\dim(G)}$ 
because $G$ is compact and so the set $U_0$ must be bounded. Furthermore, 
since the group under
consideration has only a finite number of connected components (namely one), 
there are also only a finite number of corresponding connected components of
$U_0$ and therefore the set $U-U_0$ has at most dimension
$\dim(G)-1$. It follows that $U-U_0$ has Lebesgue measure zero, that is,
we can replace the integral over $U_0$ with respect to $d^{\dim(G)}t$
by an integral over $U$. For instance, for $U(1)$ the set $U$ is just given
by the interval $[-\pi,\pi]$ while $U_0$ could be chosen as 
$[-\pi,\pi)$. Likewise, for $SU(2)$ the set $U$ is the set of points
$t_1^2+t_2^2+t_3^2\le\pi$ while $U_0$ is the set of points 
$t_1^2+t_2^2+t_3^2<\pi$ plus one arbitrary additional point of radius $\pi$ 
corresponding to the element $-1_2$.\\
Inserting (4.19),
(4.20) and (4.22) into (4.13) we can therefore write an expansion in
$1/\sqrt{\beta}$ 
\ba & & d_{\{m\}}[J_{\{m\}}(\beta)-J_0(\beta)]=\int_U
\frac{d^{\dim(G)}t}{1+o(t^2)}
\exp(-\beta\frac{1}{2}\sum_{I=1}^{\dim(G)}(t^I)^2+\beta o(t^3))\times
\nonumber\\ & & \times [t^I\mbox{tr}(X_I)+\frac{1}{2}t^I
t^J\mbox{tr}(X_I X_J)+o(t^3)]\nonumber\\ &=&
\frac{1}{\beta^{\dim(G)/2+1}}\int_{\sqrt{\beta} U}
\frac{d^{\dim(G)}t}{1+o(t^2/\beta)} \exp(-\frac{1}{2}
\sum_{I=1}^{\dim(G)}(t^I)^2+o(t^3/\sqrt{\beta}))\times\nonumber\\ & &
\times [\sqrt{\beta}t^I\mbox{tr} (X_I)+\frac{1}{2}t^I t^J\mbox{tr}(X_I
X_J)+o(t^3/\sqrt{\beta})]\nonumber\\ &=&
\frac{1}{\beta^{\dim(G)/2+1}}\int_{\Rl^{\dim(G)}} d^{\dim(G)}t
\exp(-\frac{1}{2}\sum_{I=1}^{\dim(G)}(t^I)^2)\times\nonumber\\ & &
\times[\sqrt{\beta}t^I\mbox{tr}(X_I) +\frac{1}{2}t^I t^J\mbox{tr}(X_I
X_J)+o(t^3/\sqrt{\beta})] 
\ea 
where in the last step the expansion of the scaled domain
$\sqrt{\beta}U,\; U$ a compact subset of $\Rl^{\dim(G)}$ to all of
$\Rl^{\dim(G)}$ also is correct up to a further order in
$1/\sqrt{\beta}$. Now the terms of odd order in $t$ vanish due to the
symmetry of the exponential under reflection. Therefore, we have:
\beq
d_{\{m\}}[J_{\{m\}}(\beta)-J_0(\beta)]=\frac{1}{\beta}J_0(\beta)
\frac{1}{2}\mbox{tr}(\sum_{I=1}^{\dim(G)} (X^I)^2)+o(1/\beta^2) \;.
\eeq
But $\sum_I (X_I)^2=-\lambda_{\{m\}}1_{\{m\}}$ is the Casimir
invariant and $\lambda_{\{m\}}$ is its eigenvalue. Therefore we arrive
finally at 
\beq c(N,\{m\})=\frac{1}{2}\lambda_{\{m\}} \;. \eeq 
It is well-known \cite{23} that the Laplace-Beltrami operator
$-\Delta$ has eigenvalues $\lambda_{\{m\}}$ on its complete system of
conjugation invariant eigenfunctions $\chi_{\{m\}}(g)$. These
functions are parametrized by $r$ discrete quantum numbers, according
to the rank of $G$.

\section{The Hamiltonian formalism}

In this section, we will recall the standard Hamiltonian formulation
of Lorentzian Yang-Mills theory in 1+1 dimensions. (For details, see,
e.g., \cite{9}).  Here we will only consider topologies $M = \Rl^2$
and $M = S^1 \times \Rl$ since the Lorentzian metric, obtained by
analytic continuation, on the torus $S^1\times S^1$ has closed
time-like curves. This discussion will be used in section 5.3 to show the
equivalence of our Euclidean framework with the standard Hamiltonian
description.

The canonical form of the Yang-Mills actions  is given by
\beq S=\int_R dt\int_\Sigma dx[\dot{A}_I E^I-[-\Lambda^I{\cal G}_I
+\frac{g_0^2}{2} E^I E^I]]
\eeq
where $\Sigma=\Rl\mbox{ or }S^1$ and a dot (prime) denotes a
derivative with respect to $t$ ($x$).  Here $A=A_x$ is the the
$x$-component of the G connection and $E=\frac{1}{g_0^2}(\partial_t
A_x-\partial_x A_t +[A_t,A_x])$ is its electric field. The indices
I,J,K run 1,..,dim(G) and are raised and lowered with respect to the
Cartan Killing metric. Note that time component $A_t^I = \Lambda^I$ of
the connection acts as a Lagrange multiplier, enforcing the Gauss
constraint
\beq
{\cal G}_I=E_I'+[A,E]_I\;.
\eeq
Because 
the magnetic fields vanish in one
spatial dimension, the Hamiltonian takes the form
\beq H=\int_\Sigma dx \frac{g_0^2}{2} E^I E^I. 
\eeq 
However, multiplying the Gauss constraint by $E^I$ yields
\[ \frac{1}{2}(E^I E^I)'=0 \]
so that the Hamiltonian density must be a constant.  Thus, the energy
on the plane is infinite unless that constant is zero. This enforces
the new first class constraints $E^I=0$. The motions generated by
these constraints are transitive on the whole configuration space of
the $A_I$ and so $A_I$ is identified with the trivial connection
$A_I=0$. The reduced phase space for $M = \Rl^2$ is therefore
zero-dimensional, it consists only of one point, $(0,0)$, say.

On the cylinder, the theory is less trivial and the Hamiltonian is
given by 
\beq H=\frac{g_0^2 L_x}{2}(E^I E^I). \eeq 
By a gauge transformation \cite{9}, we may take $A,E$ to be constant.
By means of a constant gauge transformation we achieve that $A$ lies
in a Cartan subalgebra. Since in that gauge the Gauss constraint
implies that $A,E$ commute, it follows that there is a gauge in which
$A,E$ both lie in a Cartan subalgebra.  Let $r$ be the rank of $L(G)$;
then the maximal Cartan subalgebra has dimension $r$ and the reduced
phase space has dimension $2r$.  The reduced phase space is then the
quotient \cite{9} of $\Rl^{2r}$ by a discrete set of residual gauge
transformations.

In the quantum theory on a cylinder, the  Hamiltonian becomes the
Laplace Beltrami operator on the Cartan subgroup $G_C$ \cite{23}
\be H=-\frac{g_0^2 L_x}{2} \Delta \ee
and physical states correspond to conjugation invariant functions on
$G$.  The corresponding inner product is the $L^2$ inner product given
by the Haar measure on $G_C$.  As a result, the characters
$\chi_{\{m\}}(A)$ with $\{m\}=\{m_1,..,m_r\}\mbox{ with } m_1\ge
m_2\ge .. \ge m_r\ge 0$ provide a complete set of eigenstates of $H$
(with eigenvalues $\frac{g_0^2 L_x\lambda_{\{m\}}}{2}$).  For
comparison with the classical theory, recall that the characters
$\chi_{\{m\}}$ depend only on the Cartan subgroup of $G$.

\section{Axiomatic framework and relation to the Hamiltonian theory}

In scalar field theories, the Osterwalder-Schrader axiomatic framework
provides a compact formulation of what is often referred to as ``the
main problem''. Consequently, the framework plays a central role in
constructive quantum field theory. However, as mentioned in the
Introduction, this framework is geared to ``kinematically linear''
theories because a basic premise of the axioms is that the space of
paths is a vector space, generally taken to be the space ${\cal S}'$
of tempered distributions. In this section, we will use the material
presented in sections 2 and 4 to suggest a possible generalization of
the Osterwalder-Schrader framework to gauge theories, using for space
of physical paths the non-linear space $\AmGb$.

The section will be divided in to three parts. In the first, we
briefly review the aspects of the Osterwalder-Schrader framework that
are relevant for our discussion. In the second, we propose an
extension of the key axioms and verify that they are satisfied by the
continuum $SU(N)$ Yang-Mills theories. In the third part we show that
the axioms suffice to demonstrate the equivalence between the
Euclidean and the Hamiltonian frameworks.  

\subsection{Kinematically linear theories}

As mentioned in the Introduction, the basic idea of the Euclidean
constructive quantum field theory \cite{11} is to {\it define} a
quantum field theory through the measure $\mu$ on the space of paths
$\Phi$ --the rigorous analog of ``$\exp -S(\Phi) {\cal D}\!\phi$''.  In
the Osterwalder-Schrader framework, the space of paths is taken to be
the space ${\cal S}'$ of tempered distributions on the Euclidean
space-time $\Rl^{d}$, and conditions on permissible measures $\mu$ on
${\cal S}'$ are formulated as axioms on their Fourier transforms
$\chi(f)$, defined via
\beq \label{5.1}
\chi(f):=<\exp(i\bar{\Phi}[f])>:=\int_{{\cal S}'}d\mu(\bar{\Phi})
\exp(i\bar{\Phi}[f])\, . 
\eeq 
Here $f$ are test functions in the Schwartz space ${\cal S}$, the
over-bar is used to emphasize that the fields are distributional and
$\bar{\Phi}[f]=\int_{\Rl^{d}} d^{d}x \bar{\Phi}(x) f(x)$ denotes the
canonical pairing between distributions and test functions. The
generating functional $\chi(f)$ determines the measure completely.
Furthermore, since ${\cal S}$ is a nuclear space, Minlos' theorem
\cite{12} ensures that if we begin with {\it any} continuous, positive
linear functional $\chi$ on ${\cal S}$, there exists a regular measure
$\mu$ on ${\cal S}'$ such that (\ref{5.1}) holds.

In the Osterwalder-Schrader framework, then, {\sl a quantum field
theory is a normalized measure $\mu$ on ${\cal S}'$, or, equivalently,
a continuous, positive linear functional $\chi$ on ${\cal S}$
satisfying the following axioms}
\begin{itemize}
\item OS-I) {\it Analyticity.} This assumption ensures that the
measure $\mu$ has an appropriate ``fall-off''. It requires that
$\chi(\sum_{i=1}^n z_i f_i)$ is entire analytic on $\Co^n$ for every
finite dimensional subspace spanned by the linearly independent
vectors $f_i\in \cal S$.\\ 
\item OS-II) {\it Regularity.} These are technical assumptions
which, roughly speaking, allow to construct Euclidean field operators
such that its Schwinger distributions
\[ S(x_1,..,x_n):=<\bar{\Phi}(x_1),..,\bar{\Phi}(x_n)> \]
are {\em tempered} --rather than less well-behaved--  distributions,
We will not display them here.\\ 
\item OS-III) {\it Euclidean Invariance.} This condition ensures
Poincar\'e invariance of the Wick-rotated theory. If $gf$ is the image
of a test function $f$ under the action of an element $g$ of the {\em
full} Euclidean group $E$ in d dimensions then, one requires:
\[ \chi(g f)=\chi(f). \] 
Here the test functions are considered as scalars, that is 
$(gf)(x):=f(gx)$.\\
\item  OS-IV) {\it Reflection positivity.} This is perhaps the key
axiom because it enables one to reformulate the theory in terms of
more familiar concepts by providing a notion of time, a Hilbert space,
and a Hamiltonian. The precise condition can be formulated as follows.
Choose an arbitrary hyper-plane in $\Rl^d$ which we will call the time
zero plane. Consider the linear space, denoted $V$, generated by finite
linear combinations of the following functions on ${\cal S}'$
\[ \Psi_{\{z_i\},\{f_i\}}\; :\; {\cal S}'\rightarrow \Co\; ;\;\quad
\bar{\Phi}\rightarrow \sum_{i=1}^n z_i\exp(\bar{\Phi}[f_i]) \]
where $z_i\in C,\; f_i\in {\cal S}$ with support only in the
``positive time'' part of the space-time (supp$(f_i)=\{x=(x^0,
\vec{x})\in \Rl^d\; ;\; x^0 >0\}$). Next, let
$\Theta(x^0,\vec{x})=(-x^0,\vec{x})$ denote the time reflection
operator ($\Theta\in E$).  Then, one requires that
\beq\label{5.2}
(\Psi,\Xi):=<\Theta\Psi,\Xi>:=\int_{{\cal S}'} d\mu(\bar{\Phi})\,\,
\overline{\Theta\Psi[\bar{\Phi}]}\, \Xi[\bar{\Phi}] \quad \ge 0\, . 
\eeq 
\item OS-V) {\it Clustering.} This axiom ensures uniqueness of the vacuum.
It requires that the measure has the cluster property, that is,
\[ \lim_{t\to\infty}\frac{1}{t}\int_0^t ds <\Psi T(s) \Xi>
=<\Psi><\Xi> \]
for all $\Psi,\Xi$ in a dense subspace of $L_2({\cal S}',d\mu)$. 
Here $T(s)$ is a representation on $L_2({\cal S}',d\mu)$ 
of the one-parameter semi-group of time translations defined by 
$\hat{T}(s)\exp(\bar{i\Phi(f)}):=\exp(i\bar{\Phi}(T(s)f)$ and
extended by linearity and $(T(s)f)(x^0,\vec{x}):=f(x^0+s,\vec{x})$
for all $f\in{\cal S}$.\\
\end{itemize}

With these axioms at hand, one can construct a Hilbert space $\cal H$
of quantum states, a Hamiltonian $H$ and a unique vacuum vector $\Omega$
(annihilated by $H$) as follows:
\begin{itemize}
\item[1)] Consider the {\em null space} $\cal N$ of norm zero vectors
in V with respect to the bilinear form ( , ) introduced in
(\ref{5.2}) and Cauchy-complete the quotient $V/\cal N$. Then ${\cal
H}:=\overline{V/{\cal N}}$ with scalar product ( , ).
\item[2)] The most important theorem now is that, given a probability
measure $\mu$ satisfying reflection positivity and Euclidean
invariance, the time translation operator $T(s)$ acting on V factors
through the quotient construction referred to in 1), that is, it leaves
the null space $\cal N$ invariant.  This means that we can represent
it on $\cal H$ and standard Hilbert space techniques now ensure that
$T(t)$ has a positive self-adjoint generator $H$ such that
$T(t)=\exp(-t H)$ (note that (due to Euclidean invariance) T(t) is
unitary with respect to $<\;,\; >$ but symmetric with respect to ( , )
due to the additional time reflection involved; this shows that T(t)
provides a symmetric contraction semi-group).
\item[3)] The vacuum state turns out to be just the projection to
${\cal H}$ of the function $1$ on ${\cal S}'$.
\end{itemize}

\subsection{A proposal for gauge theories}

Discussion of section 2 suggests that, in certain gauge theories, it is
natural to use $\AmGb$ as the space of physical paths. Thus, we are
led to seek an extension of the Osterwalder-Schrader framework in
which the linear space ${\cal S}'$ is replaced by the non-linear space
$\AmGb$. At first this goal seems very difficult to reach because the
standard framework uses the underlying linearity in almost every
step. However, we will see that one can exploit the ``non-linear
duality'' between connections and loops --or, more precisely, between
$\AmGb$ and the hoop group $\HG$-- very effectively to extend those
features of the standard framework which are essential to the proof of
equivalence between the Euclidean and the Hamiltonian frameworks.

Let us consider a gauge theory in $d$ Euclidean space-time dimensions
with a compact Lie group $G$ as the structure group. The proposal is
to use $\AmGb$ as the space of Euclidean paths. (Eventhough we are now
working in an arbitrary dimension and with more general structure
groups, this space can be again constructed using any one of the three
methods discussed in section 2.) Since $\AmGb$ is compact, it admits
normalized, regular Borel measures. Furthermore, the Riesz-Markov
theorem (together with the Gel'fand theory) ensures
\cite{3} that each of these measures $\mu$ is completely determined by
the ``characteristic functional'' $\chi(\alpha_1, ..., \alpha_n)$,
defined by:
\beq \label{5.3}
\chi(\alpha_1, ..., \alpha_n) := \int_{\AmGb} d\mu\,
\check{T}_{\alpha_1}\, ... \, \check{T}_{\alpha_n}\, , \eeq
where, $\check{T}_{\alpha_k}$ denotes Gel'fand transform of
$T_{\alpha_k}$, the trace of the holonomy around the closed loop
$\alpha_k$. (There is also a theorem \cite{it} that ensures the
converse, i.e., which states that given a functional $\chi$ of
multi-loops satisfying certain conditions, there exists a regular
measure $\mu$ on $\AmGb$ such that $\chi$ can be reconstructed via
(\ref{5.3}). However, since one has to introduce more technical
machinery to state this theorem properly and since this converse is
not logically necessary for the constructions that follow, we will not
discuss it here.) Comparing (\ref{5.3}) and (\ref{5.1}), we see that
$\AmGb$ now plays the role of ${\cal S}'$ and multi-loops, the role of
test functions, and traces of holonomies, the role of $\exp
i\bar{\Phi}(f)$. Thus, we have extended the Fourier transform
({\ref{5.1}) to a non-linear space by exploiting the fact that the
loops and connections can be regarded as ``dual objects'' in the
expression of the trace of the holonomy.  Our strategy now is to
introduce a set of axioms on measures $\mu$ through their
characteristic functionals $\chi$.
 
Let us begin with an observation. The discussion of the previous
section brings out the fact that while all five axioms are needed to
ensure that the resulting theory is complete and free of pathologies,
it is the middle two axioms --the Euclidean invariance and the
reflection positivity-- that play the central role in the
reconstruction of the Hamiltonian theory. We will therefore begin with
these axioms.

{\sl A quantum gauge field theory is a probability measure on $\AmGb$
satisfying the following axioms:}\\
$\bullet$ I) {\it Euclidean Invariance.} $\mu$ is invariant under the
full Euclidean group if the space-time topology is $\Rl^d$, and
under the full isometry group of the flat Euclidean metric
in more general context. In terms of characteristic function $\chi$,
we thus have:
\beq \chi(g\alpha)=\chi(\alpha)\, , \eeq
where $\alpha$ stands for a generic multi-loop $(\alpha_1, ...,
\alpha_n)$ and $g\alpha$ denotes the image of $\alpha$ under the
action of an isometry $g$.\\
$\bullet$ II) {\it Reflection Positivity.}  Choose, as before, an
arbitrary ``hyperplane'' and regard it as the time-zero slice. Consider
the linear space $V$ generated by finite linear combinations of
functionals on $\AmGb$ of the form
\[ \Psi_{\{z_i\},\{\alpha_{Ii}\}} \; :\; \AmGb\to \Co; \;\quad  
\bar{A}\to \sum_{i=1}^n z_i \prod_{I=1}^r 
\check{T}_{\alpha_{Ii}}(\bar{A})\, , \] 
where the loops $\alpha_{Ii}$ have support in the positive half space.
Then we must have:
\beq\label{5.4}
(\Psi,\Xi):=<\Theta\Psi,\Xi>:=\int_{\AmGb}
d\mu(\bar{A})\,\, (\Theta\Psi[\bar{A}])^\star \, \Xi[\bar{A}]
\quad \ge 0\, ,  
\eeq
where, as before $\Theta$ is the time-reflection operator.\\
$\bullet$ III) {\it Clustering.} The requirement is the same in formulae as 
for the kinematically linear field theories namely
\[ \lim_{t\to\infty}\frac{1}{t}\int_0^t ds <\Psi T(s) \Xi>
=<\Psi><\Xi> \]
for all $\Psi,\Xi$ in a dense subspace of $L_2(\AmGb,d\mu)$. Here
$T(s)T_{\alpha_1}..T_{\alpha_r}=T_{T(s)\alpha_1}..T_{T(s)\alpha_r}$
where 
$(T(s)\alpha)^0(\tau):=\alpha^0(\tau)+s,(T(s)\vec{\alpha})(\tau):=
\vec{\alpha}(\tau)$ and $\tau$ is a parameter along the loop.\\

We will see in the next section that these axioms suffice to
reconstruct the Hamiltonian theory.  However, this set of axioms is
clearly incomplete (see, e.g., \cite{13}). We will now indicate how
one might impose additional conditions and point out some subtleties.

Let us begin with the analyticity axiom of Osterwalder and Schrader.
In that case, we could take complex linear combinations $\sum z_i f_i$
because the space ${\cal S}$ of test functions is a vector space. In
the present case, we can only compose loops (or, more precisely,
hoops) to obtain
\[ \alpha =\alpha_{1}^{n_{1}}\circ ...\circ 
\alpha_{n}^{n_{n}}, \quad i=1...r \,  , \]
with integer winding numbers $n_{j}$, and, more generally, a full
subgroup of the hoop group generated by a finite number of independent
hoops (the notion of ``strong independence'' \cite{2}, of hoops being
the substitute for ``linear independence'' of test functions $f_i$.)
One could also include complex winding numbers and this may lead us
to the notion of ``extended loops'' \cite{14}. In any case, it may
be natural to require that
\[ \chi(\{\alpha_i\}) \]
be ``in some sense analytic" in the winding numbers $n_{ij}$ (we will leave
a more precise formulation of this notion for future work).
Recall, however, that in
the Osterwalder-Schrader framework, the analyticity axiom is needed to
ensure the existence of Schwinger functions. In the present case, on
the other hand, since the analogs $<A(x_1), ... , A(x_n)>$ of the
Schwinger functions fail to be gauge invariant, from our perspective,
it is unnatural to require that they be well-behaved in the quantum
theory. So, at this stage of our understanding, the ``raison d'etre''
of the analyticity condition is not as compelling in our
framework. Therefore, a definitive formulation of this axiom must
await further development of the framework.

The situation with the Regularity axiom is similar. In the Osterwalder
Schrader framework, it prescribes certain bounds on the characteristic
function $\chi(f)$ which are needed to ensure that the Schwinger
functions can be continued analytically to obtain the Wightman
functions in the Lorentzian regime. In the present context, neither
the Schwinger nor the Wightman functions are gauge
invariant. Nonetheless, suitable regularity conditions {\it are}
needed to ensure that the Lorentzian {\it Wilson loops} are
well-defined. The precise form of these conditions will become clear
only after the issue of analytic continuation of Wilson loops is
explored in greater detail. 

Finally, the space $\AmGb$ is very large: In a well-defined sense, it
serves as the ``universal home'' for measures in theories in which the
traces of holonomies are well-defined operators \cite{26}. From
general considerations, one would expect that the measures that come
from physically interesting gauge theories should have a much smaller
support (provided, of course, that traces of holonomies are measurable 
functions). A further investigation of this issue 
would suggest additional restrictions on the characteristic functions.

To conclude this section, let us consider the key question that any
set of axioms must face: Are they consistent? That is, do they admit
{\it non-trivial} examples? Fortunately, results in section 4
immediately imply that the answer is in the affirmative. To see this,
let us take $M$ to be either a 2-plane or a 2-cylinder and the structure
group to be $SU(N)$ or $U(1)$. The characteristic functional is then
given by (\ref{gf}).  Let us begin with Euclidean invariance. Since
the characteristic functionals depend only on the areas of the various
loops involved, they are invariant under all area preserving
diffeomorphisms and, in particular, under the isometry groups of the
underlying space-times. Reflection positivity is also satisfied
because, as we will see in the next sub-section, after dividing by
$\cal N$ we obtain a scalar product which is positive definite.
Furthermore, since the measure is non-interacting, clustering is 
immediate (see next subsection).
Finally, we can also test if the ``obvious'' restrictions of
analyticity and regularity are met. By inspection, the characteristic
functionals (\ref{gf}) are formally analytic in $k_I^\pm\mbox{ and
}l_i^\pm$. Since the winding numbers $n_{j}$ are linear combinations
of these, the generating functions are formally analytic in the
winding numbers as well. Finally, the generating functionals are
bounded (by $1$).
 
\subsection{Reconstruction of the Hamiltonian theory}

Let us now construct a Hilbert space, a Hamiltonian and a vacuum via
the Osterwalder Schrader algorithm \cite{11} and verify that, for
cases treated in sections 4, this description is equivalent to
the one obtained directly using Hamiltonian methods in section 5.
Since this algorithm uses, in essence, only reflection positivity, 
it is directly applicable to our formulation of gauge theories.

The first step is to construct the null space $\cal N$ in V. Let us
fix one of the $\Psi$'s considered in axiom (II). Then we have
\[ (\Psi,\Psi)=\sum_{i,j=1}^n {z}_i^\star z_j \int_{\AmGb} 
d\mu(\bar{A}) \prod_{I,J=1}^r
(\check{T}_{\Theta(\alpha_{Ii})^{-1}}[\bar{A}])\,
(\check{T}_{\alpha_{Jj}}[\bar{A}])\;, \]
where $\mu$ is the physical measure obtained by taking the continuum
limit of (\ref{gf}), and where we we have used the fact that, since
$G$ is unitary, $(\check{T}_\alpha)^\star =\check{T}_{\alpha^{-1}}$ where
$\star$ denotes complex conjugation. 

We now need to express this equation in terms of $\chi$. Let us begin
by considering the decomposition of a multi-loop $\{\alpha_1,..,
\alpha_s\},\;s\le r$. In this decomposition, it is convenient to
separate the homotopically trivial loops from the non-trivial ones.
In the case $M= \Rl\times \Rl$, there is no homotopically non-trivial
loop. On the cylinder we can choose the horizontal loop $\gamma$ at
$t=0$ as the fiducial non-trivial loop and write every homotopically
non-trivial loop $\eta$ occurring in the multi-loop $\{\alpha_1,..,
\alpha_s\}$ as $\eta=[\eta\circ\gamma^{-1}]\circ\gamma$ where the loop
in brackets is homotopically trivial. The result will be a multi-loop
$\tilde{\alpha}_1,..,\tilde{\alpha}_{\tilde{s}}$ whose homotopically
trivial contribution comes {\it only} from $\gamma$. Finally write
$\prod_{I=1}^s T_{\alpha_I}$ as a linear combination of terms of the
form (as in (\ref{prod})) 
\be
\mbox{tr}(g_{\{m\}}(\hat{\alpha}_1,..,\hat{\alpha}_{\tilde{s}})
\pi_{\{m\}}(h_\gamma)) 
\ee 
where $\pi_{\{m\}}$ is the $\{m\}$th irreducible representation of $G$
and $g_{\{m\}}$ is some matrix which depends only on the homotopically
trivial loops $\hat{\alpha}_i$ and which is projected from both sides
by $\pi_{\{m\}}(1_N)$, that is, $g_{\{m\}}\pi_{\{m\}}(1_N)
=\pi_{\{m\}}(1_N)g_{\{m\}}$.  Loops $\hat{\alpha_i}$ arise from
$\tilde{\alpha}_i$ by taking the simple loop decomposition of
$\tilde{\alpha_i}$ as in (\ref{prod}) and taking out the $\gamma$'s
and its inverses. Since every multi-loop functional can be so
expanded, it is sufficient to consider the scalar product among these
functionals which we will now write as 
\be
F_{\{m\}}(\beta):=F_{\{m\}}(\beta_1,..,\beta_s)
:=\mbox{tr}(g_{\{m\}}(\beta_1,..,\beta_s)\pi_{\{m\}}(h_\gamma)) 
\ee
where $\beta_i$ are homotopically trivial and enclose surfaces in the
positive half-space. Note that
$\Theta F_{\{m\}}(\beta)=F_{\{m\}}(\Theta \beta)$ since 
$\Theta\gamma=\gamma$. We can therefore alternatively write $\Psi$
in the form
\be \Psi=\sum_{\{m\}} z_{\{m\}} F_{\{m\}}(\beta_{\{m\}}) \;. \ee
Now, using the formula \cite{21}
\be \int_G d\mu_H(g) 
\bar{\pi}_{AB}(g)\otimes\pi'_{CD}(g)=\frac{\delta_{\pi,\pi'}}{d_\pi}
\pi_{AC}(1)\pi_{BD}(1)
\ee
we find
\ba\label{6.10}
& & \int_{\AmGb} d\mu \bar{F}_{\{m\}}(\Theta\beta_{\{m\}})
F_{\{m'\}}(\beta_{\{m'\}})
\nonumber\\
&=&\frac{\delta_{\{m\},\{m'\}}}{d_{\{m\}}} \int_{\AmGb} d\mu 
\bar{g}_{\{m\}}(\Theta\beta_{\{m\}})_{AB} 
g_{\{m\}}(\beta_{\{m\}}))_{BA}
\nonumber\\ 
&=&\frac{\delta_{\{m\},\{m'\}}}{d_{\{m\}}}\mbox{tr}
([\int_{\AmGb} d\mu \bar{g}_{\{m\}}(\beta_{\{m\}})] 
[\int_{\AmGb} d\mu g_{\{m\}}(\beta_{\{m\}})])
\nonumber\\
&=&\frac{\delta_{\{m\},\{m'\}}}{d_{\{m\}}^2}
|\int_{\AmGb} d\mu\; 
\mbox{tr}(\pi_{\{m\}}(1_N)g_{\{m\}}(\beta_{\{m\}}))|^2 \;.
\ea
Here, in the third step we have used the fact that $\beta_{\{m\}},
\Theta\beta_{\{m'\}}$ are supported in disjoint domains of space-time,
the time reflection invariance of the measure and its maximal
clustering property of the measure (non-overlapping loops are
non-interacting). In the last step we used the fact that the integral
over $g_{\{m\}}(\beta_{\{m\}})_{AB}$ results in a constant matrix,
$M_{AB}$ say, which, by inspection of (\ref{3.10}) is a linear
combination of projectors onto representation spaces of irreducible
representations, partially contracted as to match the index structure
of $\pi_{\{m\}}$.  So $M$ is a linear combination of matrices of the
form $\sigma'_{AB}=\sigma_{C,A;C,B}(1_N)$ where $\sigma$ is an
irreducible projector. Now using the fact that $\sigma'
\pi_{\{m\}}(1_N)=\pi_{\{m\}}(1_N)\sigma'$, that $\pi_{\{m\}}$ is
irreducible and that the contraction of tensor products of Kroneckers
is again proportional to a tensor product of Kroneckers it follows
that $M=\pi_{\{m\}}(1_N)\mbox{tr}(M)/d_{\{m\}}$.\\

Formula (\ref{6.10}) says that 
\be \label{6.11}
\Psi-\sum_{\{m\}} z_{\{m\}}\frac{1}{d_{\{m\}}}[\int_{\AmGb} 
d\mu \mbox{tr}(\pi_{\{m\}}(1_N)g_{\{m\}}(\beta_{\{m\}}))] 
\chi_{\{m\}}(h_\gamma)
\ee
is a null vector. Therefore, our Hilbert space $\cal H$ is the
completion of the linear span of the states $\chi_{\{m\}}(h_\gamma)$
with respect to the Haar measure $d\mu_H$. On the plane, since there
is no homotopically non-trivial loop $\gamma$, the only state is the
constant function $\Psi= 1$ which corresponds precisely to the trivial
quantum theory as obtained via the Hamiltonian formalism.  On the
cylinder we obtain ${\cal H}=L_2(C(G),d\tilde{\mu}_H)$ where $C(G)$ is
the Cartan subgroup of $G$ and $\tilde{\mu}_H$ is the corresponding
effective measure on $C(G)$ induced by the Haar measure $\mu_H$. 

Finally, note that, in the final picture, the loop $\gamma$ probes the
connection $\bar{A}$ at time $t=0$ only. This is is completely analogous
to the corresponding construction for the free massless scalar field
(\cite{11}) where the Hilbert space construction can be reduced to the
fields at time zero.

Having constructed the Hilbert space, let us now turn to the
Hamiltonian. As indicated in section 6.1, the Hamiltonian can be
obtained as the generator of the Euclidean time translation
semi-group. Denote by $\gamma(t):=T(t)\gamma$ the horizontal loop at
time $t$. Now let $\alpha(t):=\gamma(t)\circ\gamma^{-1}$, then we have
by the representation property
\be
\chi_{\{m\}}(h_{\gamma(t)})=\mbox{tr}(\pi_{\{m\}}(h_{\alpha(t)})
\pi_{\{m\}}(h_\gamma)) \ee so that according to (4.16) we have that
\be \chi_{\{m\}}(h_{\gamma(t)}) =[\int_{\AmGb} d\mu
\chi_{\{m\}}(\alpha(t))]\frac{\chi_{\{m\}}(h_\gamma)}{d_{\{m\}}}
\;. 
\ee 
Hence, according to (4.14)
\[ 
(\chi_{\{m'\}},T(t)\chi_{\{m\}})\, = \, \exp(-\frac{1}{2}
\lambda_{\{m\}} 
g_0^2 L_x t)\delta_{\{m'\},\{m\}}\, \stackrel{!}{=}\, 
(\chi_{\{m'\}}, \exp(-t H)\chi_{\{m\}}) 
\]  
and the completeness of the $\chi_{\{m\}}$ allows us to conclude that
\beq 
H=-\frac{g_0^2}{2}L_x\Delta 
\eeq 
is the configuration representation of the Hamiltonian.

Finally, let us consider the vacuum state. By inspection, it is given
by $\Omega=1$. It is the unique vector annihilated by the Hamiltonian.
We therefore expect that the measure is clustering 
\cite{11}, Theorem 19.7.1. Indeed, notice first that finite linear 
combinations of products of traces of the holonomy around loops form
a dense set $\cal D$ in $L_2(\AmGb,d\mu)$ by construction of $\AmGb$. Now 
recall once 
again that the measure is not interacting in the sense that if $\Psi,\Xi$
are two elements of $\cal D$ defined through multi-loops lying in disjoint 
regions of the plane or the cylinder then it follows immediately from (4.16) 
that $<\Psi\Xi>=<\Psi><\Xi>$. Even if $\Psi,\Xi$ are defined through 
multi-loops which intersect or overlap then there exists a time parameter
$t_0$ such that the multiloops involved in $\Psi$ and $T(t)\Xi$ lie
in disjoint regions of the plane or the cylinder for all $t\ge t_0$.
It then follows from the invariance of the measure under time 
translations that for $t>t_0$ we have 
\[ \int_0^t ds <\Psi T(s)\Xi>=\int_0^{t_0} ds <\Psi T(s)\Xi>
+<\Psi><\Xi>(t-t_0) \]
and since the first term is finite, clustering is immediate.

Thus, as in scalar field theories, Euclidean invariance and reflection
positivity have enabled us to construct the Hamiltonian description from
the Euclidean. Furthermore, from sections 4 and 5 it follows that for
$SU(N)$ and $U(1)$ Yang-Mills theories on $\Rl\times R1$ and
$S^1\times R1$, the Hamiltonian theory constructed through this procedure
is {\it exactly} the same as the standard one, constructed ab-initio
via canonical quantization.

\section{Summary}

The new results of the present paper can be summarized as follows:

$\bullet$ We successfully employed the new integration techniques
developed in \cite{2, 29, baez} to compute a closed expression for the
Wilson loop functionals for Yang-Mills theory in two Euclidean
dimensions.

$\bullet$ We proposed an extension of the Osterwalder-Schrader
framework for gauge theories and showed how to recover the Hilbert
space, the Hamiltonian and the vacuum for the Lorentzian theory
starting from our Euclidean framework. For 2-dimensional Yang-Mills
theories on $\Rl\times\Rl$ and on $S^1\times \Rl$, the resulting
quantum theory completely agrees with the one obtained via canonical
quantization. Therefore, two-dimensional Yang-Mills theory constitutes
another model theory in the framework of constructive quantum field
theory.

$\bullet$ Our results are manifestly gauge-invariant, geometrically
motivated, require only simple mathematical techniques and the
resulting quantum theory is manifestly invariant under the classical
symmetry generated by area-preserving diffeomorphisms.

How do these results compare with those available in the literature?
Let us begin with the Makeenko-Migdal approach. While they formulated
differential equations that the Wilson loops have to satisfy, we have
derived a general expression for Wilson loops themselves by directly
computing the functional integrals. In the intermediate steps we used
a lattice regularization. However, in contrast to the more common
practice (in lattice gauge theories) of seeking fixed points of the
renormalization group, our results for the continuum theories were
then obtained by explicitly taking the limits to remove the
regulators. Indeed, our general procedure is rather similar to that
used in constructive quantum field theory: we began with a fiducial
measure $\mu_0$ on our space $\AmGb$ of Euclidean paths, introduced an
infra-red and an ultra-violet cutoff, evaluated the characteristic
functional of the measure and then removed the regulators. Thus,
in the end, we were able to show rigorously that the theory exists in
the continuum. In particular, our mathematical framework guarantees the
existence of the physical measure for the continuum theory (for which the
``fixed point'' arguments of numerical lattice theory do not suffice.)

While the spirit of our approach is the same as that of the
mathematical physics literature on the subject, there are some
differences as well. Most of these approaches mimic techniques that
have been successful in scalar field theories. Thus, generally, one
fixes gauge right in the beginning to introduce a vector space
structure on $\AmG$ (see, e.g. \cite{1}). Proofs of invariance of the
final expressions under gauge transformations and area preserving
diffeomorphisms are then often long. Also, in most of this literature,
the Wilson loops are computed for non-overlapping loops. Our results
are perhaps closest to those of Klimek and Kondracki \cite{25}.  Their
framework is also manifestly invariant under gauge transformations and
area preserving diffeomorphisms. Furthermore, their results (as well
as those of the second paper in Ref. \cite{1}) imply that their
expressions of Wilson loops in the non-overlapping case admit
consistent extensions to all loops. However, they restrict themselves
to the structure group $SU(2)$ and the relation to lattice gauge
theory --and hence to the conventional Yang-Mills theory-- is somewhat
obscure.

There are several directions in which our results can be extended. We
will conclude by mentioning some examples. First, now that closed
expressions for Wilson loops are available, it would be very
interesting to check if they satisfy the Makeenko-Migdal equations
rigorously. Second, our axiomatic framework is very incomplete and it
would be very desirable to supplement it, e.g., with techniques from
\cite{13}. Another direction is suggested by the fact that, for
theories discussed here in detail, we expect that the support of the
final physical measure is significantly smaller than the full space
$\AmGb$ with which we began. Rigorous results that provide a good
control on the support would be very useful in refining our axiomatic
framework.  Finally, it would be interesting to extend our Euclidean
methods to closed topologies and compare the resulting framework with
the gauge fixed framework of Sengupta's \cite{sg}.  \\

{\large Acknowledgements}\\ 

Most of this research was carried out at the Center for Gravitational
Physics and Geometry at The Pennsylvania State University and JL and
JM would like to thank the Center for its warm hospitality. The
authors were supported in part by the NSF Grant PHY93-96246 and the
Eberly research fund of The Pennsylvania State University. DM was
supported in part also by the NSF Grant PHY90-08502. JL was supported
in part also by NSF Grant PHY91-07007, the Polish KBN Grant 2-P302
11207 and by research funds provided by the Erwin Schr\"odinger
Institute at Vienna. JM was supported in part also by 
research funds provided by Junta Nacional de
Investiga\c c\~ao Cientifica e Tecnologica, CERN/P/FAE/1030/95.
TT was supported in part also by DOE-Grant DE-FG02-94ER25228 to Harvard 
University.

\begin{appendix}

\section{Young tableaux}

The relevant reference for the sequel is \cite{17}.

In the main text we encounter the following problem: We have to integrate
a tensor product of group factors $\otimes^n g$ with a measure 
$d\mu=d\mu_H(g)\exp(\beta/N\Re \mbox{tr}(g))$ which is invariant under 
conjugation. The representation of G corresponding to the n-fold tensor 
product of the fundamental representation is not irreducible, 
so let us decompose it into irreducibles
\[ \otimes^n g=\oplus_i \pi^{(n)}_i(g) \]
which is possible since $G$ is compact. Now we have that 
\[ \pi(h)[\int_G d\mu(g) \pi(g)]=[\int_G d\mu(g) \pi(hgh^{-1})] \pi(h)
=[\int_G d\mu(g) \pi(g)] \pi(h) \]
so the integral over $\pi(g)$ commutes with the representation (we
have used conjugation invariance of the measure in the last
step). Accordingly, by Schur's lemma, we conclude that the integral is
proportional to the identity since $\pi$ was supposed to be an
irreducible representation. We can compute the constant of
proportionality by taking the trace. Therefore we conclude that 
\beq
\int_G d\mu(g) \pi(g)=\frac{\pi(1_N)}{d(\pi)}\int_G d\mu(g) \chi(g),
\mbox{ where }\chi(g)=\mbox{tr}(\pi(g)) 
\eeq 
is the character of the representation. This simplifies the group
integrals significantly since we only need the character integrals.

Note that what we are doing here is different from what is usually
done in the literature \cite{8,18}: Because we want to evaluate the
integral non-perturbatively, we cannot use the stronger property
of translation invariance of the Haar measure. In case of the Haar
measure we simply have \cite{8} 
\be \int_G d\mu_H(g)\pi(g)=\delta_{\pi,0}\pi(1_N) \ee 
where $0$ denotes the
trivial representation.

The solution to the problem of how to decompose an arbitrary tensor
product of fundamental representations of $SU(N)$ into irreducibles
can be found, e.g., in \cite{17} and we just recall the necessary
parts of the theory.

Given an n-fold tensor product of the fundamental representation of a
group G, consider all possible partitions $\{m\}$ of n into positive
integers of decreasing value,
\[ n=m_1+m_2+..+m_s,\mbox{ where }m_1\ge m_2\ge ..\ge m_s>0\; . \]
Such a partition defines a so-called {\em frame} Y (Young diagram) 
composed of s horizontal rows with $m_i$ boxes in the i-th row. 

Associated with each frame we construct a certain operator acting on
the n-fold tensor product representation as follows : Fill the boxes
arbitrarily with numbers $B_1, B_2, ...,B_n$ where $B_i\in\{1,2,..,N\}$.
Such a filling of the frame is called a {\em tableau}.  Let P denote
the subset of the symmetric group of n elements $S_n$ which only
permutes the indices i of the labels $B_i$ of each row among
themselves and similarly Q denotes the subgroup of $S_n$ permuting
only the indices in each column among themselves of the given
frame. The relevant operator is now given by
\[ e^{(n)}_{\{m\},i}:=\sum_{q\in Q}\mbox{sgn}(q) \hat{q}
\sum_{p\in P} \hat{p} \]
where i labels the filling and sgn(q) denotes the sign of the
permutation q. The action of $\hat{p}$, say, is
\[ \hat{p} g^{A_1}_{B_1}... 
g^{A_n}_{B_n}=g^{A_1}_{B_{p(1)}}..g^{A_n}_{B_{p(n)}}, \] 
that is, it permutes the {\em indices} of the {\em subscript} labels
$B_i$. Because of the complete anti-symmetrization in the columns, no
diagram has a row longer than N boxes, $s\le N$.

It turns out \cite{17} that each of these {\em symmetrizers}
corresponds to an irreducible representation of $GL(N),U(N)$ and $SU(N)$. 
Symmetrizers
corresponding to different frames give rise to inequivalent
representations all of those that correspond to different fillings of
the same frame are equivalent.  However, not all of the symmetrizers
for a given frame are linearly independent, a linearly independent set
of tableaux, the so-called {\em standard tableaux} can be constructed
as follows : let the indices i of a filling always increase in one row
from left to right and in each column from top to bottom. The number
of these standard tableaux is given by the formula (if $s=1$, replace
the numerator of the fraction by 1) 
\beq
f^{(n)}_{\{m\}}:=n!\frac{\prod_{1\le i<j\le s}(l_i-l_j)}
{\prod_{i=1}^s (l_i !)}\mbox{ where }l_i:=m_i+s-i,\;i=1,..,s 
\eeq
and it is the number of times that the $\{m\}$th irreducible
representation occurs in the decomposition of $\otimes^n g$ into
irreducibles.\\ Now let 
\beq
e_{\{m\}}^{(n)}:=\sum_{i=1}^{f^{(n)}_{\{m\}}} e^{(n)}_{\{m\},i} 
\eeq
i.e. the sum of the symmetrizers corresponding to the standard
tableaux.  This object is called the {\em Young symmetrizer} of the
frame $\{m\}$.  One can show that the standard symmetrizers obey the
following (quasi) projector property
\[ [e^{(n)}_{\{m\},i}\otimes^n 1_N][e^{(n)}_{\{m'\},j}
\otimes^n 1_N] =\delta_{i,j}\delta(\{m\},\{m'\})\frac{n!}
{f^{(n)}_{\{m\}}} e^{(n)}_{\{m\},i}\;, \]
that is, the sum in (A.4) is actually direct and 
\[ p^{(n)}_{\{m\},i}:=\frac{f^{(n)}_{\{m\}}}{n!}e^{(n)}_{\{m\},i}
\mbox{ and }
 p^{(n)}_{\{m\}}:=\frac{f^{(n)}_{\{m\}}}{n!}e^{(n)}_{\{m\}} \] 
are projectors onto the representation space of the i-th of the
equivalent irreducible standard representations given by the frame and
on their direct sum respectively.\\ In particular we have the
resolution of the identity 
\beq 
\otimes^n 1_N=\oplus_{\{m\}}
[p^{(n)}_{\{m\}}\otimes^n 1_N]. 
\eeq 
Let us focus on the unitary groups from now on. For the groups SU(N)
we have the following formula for the dimension of the $\{m\}$-th
irreducible representation \cite{18} : 
\beq
d_{\{m\}}=\frac{\prod_{1\le i<j\le N}(k_i-k_j)}{\prod_{I=1}^{N-1} (I
!)} \mbox{ where } k_i=m_i+N-i,\;m_i:=0\mbox{ for }i>s. 
\eeq

\section{U(1) on the torus}

According to the formulas developed in sections 3 and 4.1 
it is easy to see that the characteristic functional simply becomes
\ba & & \chi(\alpha)=\frac{\int\prod_{\Box}d\mu_H(g_\Box) d\mu_H(g_x)
d\mu_H(g_y)
\exp(-\beta\sum_\Box[1-\Re(g_\Box)])T_\alpha(g_\Box,g_x,g_y)
\delta(\prod_\Box g_\Box,1)}{\int\prod_\Box
d\mu_H(g_\Box)\exp(-\beta\sum_\Box[1-\Re(g_\Box)]) \delta(\prod_\Box
g_\Box,1)}\nonumber\\ &=&\delta_{l_x,0}\delta_{l_y,0}\lim_{N\to\infty}
\frac{\sum_{n=-N}^N (\frac{I_n}{I_0})^{N_x N_y}\prod_{I=1}^k
(\frac{I_{n+k_I}/I_0}{I_n/I_0})^{|\alpha_I|}} {\sum_{n=-N}^N
(\frac{I_n}{I_0})^{N_x N_y}} \ea
where we have employed in the second step the Dirichlet formula
\cite{22}
\be \delta(g,1)=\sum_{n=-\infty}^\infty g^n \ee
and we could interchange the processes of taking the limit and
integration since the Wilson action satisfies all the regularity
assumptions for the application of that
formula. $I_n(\beta)=\int_{-\pi}^\pi d\phi/(2\pi)
e^{\beta\cos(\phi)+in\phi}$ is the $n-th$ modified Bessel function.

Let us write $N_x N_y=\beta g_0^2 V\mbox{ and }|\alpha_I|=\beta g_0^2
A(\alpha_I)$ ($V$ is the volume or total area of the torus and
$A_I=A(\alpha_I)$ are the areas of the simple non-overlapping
homotopically trivial loops of which $\alpha$ is composed) and use the
well-known asymptotic properties of the modified Bessel functions
\cite{fl} in taking the continuum limit $\beta\to\infty$. The result
is 
\ba 
& & \chi(\alpha)=\delta_{l_x,0}\delta_{l_y,0}\lim_{N\to\infty}
\frac{\sum_{n=-N}^N \exp(-\frac{g_0^2}{2}[n^2 V-\sum_{I=1}^k
A_I([n+k_I]^2-n^2)])} {\sum_{n=-N}^N\exp(-\frac{g_0^2}{2}V n^2)}
\nonumber\\ & =
&\delta_{l_x,0}\delta_{l_y,0}e^{-\frac{g_0^2}{2}[\sum_{I=1}^k A_I
k_I^2 -\frac{1}{V}(\sum_{I=1}^k k_I A_I)^2]}
\frac{\sum_{n=-\infty}^\infty\exp(-\frac{V
g_0^2}{2}(n+\frac{\sum_{I=1}^k A_I k_I}{V})^2)}
{\sum_{n=-\infty}^\infty \exp(-\frac{V g_0^2}{2}n^2)}\;.  
\ea
Note that the series in numerator and denominator converge absolutely
and uniformly to a non-vanishing limit.

Formula (B.3) is the exact and complete result. If we could replace
the sums by integrals over the real axis then the fraction involved in
(C.3) would give just the number 1 and we would be left with the
exponential factor only. Note that because
$V-A_I\ge\sum_{J\not=I}A_J$, exponent of the exponential is negative :
\be V\sum_I A_I k_I^2-(\sum_I k_I A_I)^2\ge\sum_{I,J\not=I} k_I^2 A_I
A_J -2\sum_{I<J}k_I k_J A_I A_J=\sum_{I<J} A_I A_J(k_I-k_J)^2\ge 0 \ee
so that this pre-factor alone could possibly be the generating
functional of a positive measure (According to the Riesz-Markov
theorem one needed to verify that it is a positive linear functional
on $\cal HA$).

The characteristic functional (B.3) has several interesting features, for
example :\\ 1) While the non-interacting measures had exponents that
were linear in the areas of the simple loops, for the interacting
theory on the torus we obtain a quadratic dependence on the area, thus
{\em violating the area law} ! It is an interesting speculation that
the interactive nature of the measure is related to the fact that
functional integrals with compact time direction are supposed to
describe finite temperature field theories. The interaction then comes
from the background heat bath and the characteristic functional is the
free energy of a canonical ensemble.\\ 2) It is invariant under taking
complements (that is, $A\to V-A$) if there is only one simple loop,
otherwise the simple loop decomposition of the complemented surfaces
is different from the original one.
\end{appendix}

\end{document}